\documentclass[aps,prl,floatfix,twocolumn,showpacs,nofitinbib,hyperref=pdftex,citeautoscript]{revtex4-1}
\usepackage{epsfig}
\usepackage{amsmath}
\usepackage{amssymb}
\usepackage{amsfonts}
\usepackage{ dsfont }
\usepackage{ comment,color }
\usepackage{lipsum}
\usepackage{hyperref}

\hypersetup{colorlinks=true,linkcolor=blue,anchorcolor=blue,citecolor=blue,filecolor=blue,urlcolor=blue,bookmarksnumbered=true,pdfview=FitB}

\newcommand{\bk}{{{\bf{k}}}}

\newcommand{\bq}{{\bf{q}}}

\newcommand{\beqa}{\begin{eqnarray}}
\newcommand{\eeqa}{\end{eqnarray}}

\newcommand{\ua}{\uparrow}
\newcommand{\da}{\downarrow}

\begin{document}

\hsize\textwidth\columnwidth\hsize\csname@twocolumnfalse\endcsname

\title{Quantum Computing with Majorana Kramers Pairs}
\author{Constantin Schrade and Liang Fu}

\affiliation{Department of Physics, Massachusetts Institute of Technology, 77 Massachusetts Ave., Cambridge, MA 02139}

\date{\today}

\vskip1.5truecm
\begin{abstract}
We propose a universal gate set acting on a qubit formed by the degenerate ground states of a Coulomb-blockaded time-reversal invariant topological superconductor island with spatially separated Majorana Kramers pairs: the ``Majorana Kramers Qubit". All gate operations are implemented by coupling the Majorana Kramers pairs to conventional superconducting leads. 
Interestingly, in such an all-superconducting device, the energy gap of the leads provides another layer of protection from quasiparticle poisoning independent of the island charging energy. Moreover, the absence of strong magnetic fields -- which typically reduce the superconducting gap size of the island -- suggests a unique robustness of our qubit to quasiparticle poisoning due to thermal excitations. Consequently, the Majorana Kramers Qubit should benefit from prolonged coherence times and may provide an alternative route to a Majorana-based quantum computer.
 \end{abstract}

% 03.67.Lx: Quantum computation architectures and implementations
% 74.50.+r: Tunneling phenomena; Josephson effects 
% 85.25.Cp Josephson devices
% 71.10.Pm: Fermions in reduced dimensions

\maketitle
In recent years an increasing number of platforms have been proposed for realizing time-reversal invariant topological superconductors (TRI TSCs) \cite{bib:Schnyder2008}. Among the most notable platforms are nanowires and topological insulators in contact to unconventional superconductors (SCs) \cite{bib:Wong2012,bib:Nagaosa2013,bib:Zhang2013,bib:Dumitrescu2014} and conventional SCs \cite{bib:Klinovaja2014,bib:Gaidamauskas2014,bib:Schrade2017,bib:Klinovaja20142,bib:Yan2018,bib:Hsu2018}, proximity-induced Josephson $\pi$-junctions in nanowires and topological insulators \cite{bib:Keselman2013,bib:Haim2014,bib:Schrade2015,bib:Borla2017} as well as TSCs with an emergent time-reversal symmetry (TRS) \cite{bib:Huang2017,bib:Reeg2017,bib:Hu2017,bib:Maisberger2017}. 

A common feature of TRI TSCs  is that they host spatially separated Majorana Kramers pairs (MKPs) which form robust, zero energy modes protected by TRS. In spite of much fundamental interest in the properties of MKPs \cite{bib:Chung2013,bib:Li2016,bib:Pikulin2016,bib:Kim2016,bib:Camjayi2017,bib:Bao2017,bib:Schrade2018}, a yet unsolved question is if MKPs can be employed for applications in quantum computation. Here, we answer this question in the affirmative. 

\begin{figure}[!t] \centering
\includegraphics[width=0.75\linewidth] {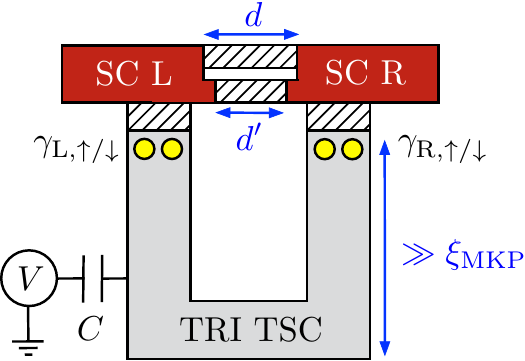}
\caption{(Color online)
Setup consisting of a U-shaped, mesoscopic TRI TSC island (gray) realizing a MKQ. Tunable tunnel couplings (white, dashed) connect SC leads $\ell=\text{L,R}$ (red) to the MKPs $\gamma_{\ell,s}$ (yellow) with $s=\ua,\da$. The SC leads themselves are also connected by a spin-flip and a normal tunnelling barrier with lengths $d, d'$. To facilitate Cooper pair splitting between these two tunnelling barriers and the TRI TSC island we require that the separation of the tunnelling contacts is smaller than the coherence length $\xi_{\text{SC}}$ of the SC leads. Moreover, to avoid couplings of the MKPs to fermionic corner modes \cite{bib:Loss2015}, the length of the vertical segments of the TRI TSC islands are much longer than the MKP localization length $\xi_{\text{MKP}}$. Lastly, a gate voltage $V$ tunes the charge on the TRI TSC island via a capacitor with capacitance $C$. 
}\label{fig:1}
\end{figure}

The purpose of this work is to introduce a qubit formed by the degenerate ground states of a Coulomb-blockaded TRI TSC island with spatially separated MKPs: the ``Majorana Kramers Qubit" (MKQ). We depict the minimal experimental setup for a single MKQ in Fig.~\ref{fig:1}. It comprises two SC leads which separately couple to two distinct MKPs on a U-shaped TRI TSC island. The two SC leads are weakly coupled among themselves by spin-flip and normal tunnelling barriers. Within this setup, we will implement single-qubit Clifford gates by making use of a measurement-based approach to quantum computing \cite{bib:Bonderson2008,bib:Litinski2017}. Moreover, to achieve universal quantum computation we will implement a $\pi/8$-gate as well as a two-MKQ entangling gate by pulsing of tunnel couplings.

The main conceptual lesson we will learn is that Majorana-based quantum computing is possible without invoking the need for magnetic fields. Besides that, there two interesting, yet more practical, features of our setup which are noteworthy: (1) Within the single-MKQ setup of Fig.~\ref{fig:1}, single-electron tunnelling from the SC leads does not only require overcoming the charging energy of the TRI TSC island but also the breaking of a Cooper pair in the leads. Consequently, the SC gap of the leads provides an additional layer of protection against quasiparticle poisoning, independent of the island charging energy. (2) Quasiparticle poisoning due to thermal excitations within the TRI TSC island is strongly suppressed the SC gap of the island itself.  Critically, the energy gap of a TRI TSC island is conceivably larger than the energy gap of TRS-breaking Majorana islands \cite{bib:Fu2010,bib:Vijay2015,bib:Vijay2016,bib:Landau2016,bib:Plugge2016,bib:Vijay2016_2,bib:Aasen2016,bib:Karzig2016,bib:Plugge2017,bib:Schrade2018_2,bib:Gau2018} since there is no magnetic field that would reduce the SC gap size. As a consequence, the MKQ should benefit from improved coherence times and may be a viable route towards a robust quantum computer.

{\it Setup.} 
As shown in Fig.~\ref{fig:1}, our setup comprises a U-shaped TRI TSC islands hosting MKPs $\gamma_{\ell,s}$ with $s=\ua,\da$ at spatially well separated boundaries $\ell=\text{L,R}$. The two members of a MKP are related by TRS,
\begin{equation}
\mathcal{T}\gamma_{\ell,\ua}\mathcal{T}^{-1}=\gamma_{\ell,\da}, \;  \mathcal{T}\gamma_{\ell,\da}\mathcal{T}^{-1}=-\gamma_{\ell,\ua}.
\end{equation}
We assume that the dimensions of the horizontal island segments exceed the localization lengths $\xi_{\text{MKP}}$ of the MKPs. This avoids couplings of the MKPs to fermionic modes that are potentially localized at the island corners \cite{bib:Loss2015} and, thereby, ensures that the MKPs are, in fact, robust zero-energy states protected by TRS.  

Since the TRI TSC island is of mesoscopic size, it acquires a charging energy given by
\begin{equation}
U_{C} = \left(ne-Q\right)^{2}/ 2C.
\end{equation}
Here, $Q$ is the island gate charges that is continuously tunable with a voltage across a capacitor with capacitance $C$. We assume that the gate charge $Q/e$ is tuned close to an even or odd integer for both islands. A sufficiently large charging energy $e^{2}/2C$ then fixes the joint parity of the MKPs on the TRI TSC island to \cite{bib:Fu2010,bib:Xu2010}
\begin{equation}
\label{TotalParity}
\gamma_{\text{L},\ua}\gamma_{\text{R},\ua}\gamma_{\text{L},\da}\gamma_{\text{R},\da} = (-1)^{n_0}.
\end{equation}
This constraint reduces the four-fold degeneracy of the ground state at zero charging energy,
to a two-fold degenerate ground state which forms the MKQ. The Pauli operators acting on each of the two MKQs 
can be written as bilinears in the Majorana operators,
\begin{equation}
\begin{split}
\hat{x}&=i\gamma_{\text{R},\ua}\gamma_{\text{L},\da}, \quad
\hat{y}=i\gamma_{\text{R},\ua}\gamma_{\text{R},\da} , \quad
\hat{z}=i\gamma_{\text{R},\da}\gamma_{\text{L},\da}.
\end{split}
\end{equation}
Under TRS, the Pauli operators transform as $\mathcal{T}\hat{x}\mathcal{T}^{-1}=(-1)^{n_0}\hat{x}$, $\mathcal{T}\hat{y}\mathcal{T}^{-1}=-\hat{y}$ and $\mathcal{T}\hat{z}\mathcal{T}^{-1}=(-1)^{n_0}\hat{z}$.

In our setup, we choose to address the MKQ by weakly coupling each MKP to a separate $s$-wave SC lead. The Hamiltonian for the two SC leads reads 
\begin{equation}
H_{SC}=\sum_{\ell=\text{L,R}}\sum_{\bk} \Psi_{\ell,\bk}^\dagger \left(
\xi_{\bk}\eta_{z}+\Delta_{\ell}\eta_{x}e^{i\varphi_{\ell}\eta_{z}}
\right)\Psi_{\ell,\bk},
\end{equation}
where $\Psi_{\ell,\bk}=(c_{\ell,\bk\ua},c^{\dag}_{\ell,-\bk\da})^{T}$ is a Nambu spinor with $c_{\ell,\bk s}$ the electron annihilation operator at momentum $\bk$ and spin $s$ in lead $\ell$. The Pauli matrices $\eta_{x,y,z}$ are acting in Nambu-space. Furthermore, $\xi_{\bk}$ is the normal state dispersion and $\Delta_{\ell},\varphi\equiv\varphi_{\text{L}}-\varphi_{\text{R}}$ denote the magnitude and the relative phase difference of the SC order parameters. We assume sufficiently low temperatures such that no quasiparticle states in the SC leads are occupied with notable probability and can couple to the MKPs. 

The most general tunneling Hamiltonian between the MKPs on the islands and the fermions on the $\ell$-SC lead is given by,
\begin{align}
\label{Eq4}
H_{T}
&=
\sum_{\ell=\text{L,R}}
\sum_{\bk,s}
\lambda_{\ell}
c^{\dag}_{\ell,\bk s}
\gamma_{\ell,s}
e^{-i\phi/2}
+
\text{H.c.}, 
\end{align}
where we have diagonalized the tunnelling Hamiltonian in spin-space by an appropriate rotation of the lead fermions \cite{bib:Schrade2018}. In particular, this rotation also constraints the point-like tunnelling amplitudes $\lambda_{\ell}$ to be real numbers. Furthermore, we remark that the operators $e^{\pm i\phi/2}$ raise/lower the total island charges by one unit, $[n,e^{\pm i\phi/2}]=\pm e^{\pm i\phi/2}$, while the MBSs operators $\gamma_{\ell,s}$ flip the respective electron number parities.

As evident from Fig.~\ref{fig:1}, there are two types of couplings between the SC leads: The first type is an \textit{indirect coupling} via the TRI TSC islands which is induced by the tunnelling Hamiltonian of Eq.~\eqref{Eq4}. The second type is a \textit{direct coupling} via two additional tunnelling barriers. The first tunnelling barrier only allows for normal tunnelling described by the tunnelling Hamiltonian, 
 \begin{equation}
H_{N}
=
t_{N}
\sum_{\bk}
c^{\dag}_{\text{R},\bk \ua}c_{\text{L},\bk \ua} 
+
c^{\dag}_{\text{L},\bk \da} c_{\text{R},\bk \da}
+ \text{H.c.}, 
\end{equation}
with $t_{N}$ a complex, point-like tunnelling amplitude and $|t_{N}|\ll\lambda_{\ell}$. The second barrier only allows for spin-flip tunnelling described by a tunnelling Hamiltonian.
\begin{equation}
H_{S}
=
t_{S}
\sum_{\bk}
c^{\dag}_{\text{R},\bk \ua}c_{\text{L},\bk \da} 
-
c^{\dag}_{\text{L},\bk \ua} c_{\text{R},\bk \da}
+ \text{H.c.},
\end{equation}
with $t_{S}\ll\lambda_{\ell}$ a complex, point-like tunnelling amplitude and $|t_{S}|\ll\lambda_{\ell}$. We propose two ways to engineer such tunnellling barriers: (1) We consider barriers with a finite \textit{intrinsic} spin-orbit coupling with spin-orbit length $\lambda_{\text{SO}}$ as well as different barrier lengths $d,d'$. Tuning $\lambda_{\text{SO}}/d'$ ($\lambda_{\text{SO}}/d$) to a positive integer (positive half integer) realizes a barrier with pure normal (spin-flip) tunnelling \cite{bib:Bercioux2015}. (2) We consider barriers without intrinsic spin-orbit coupling but with an \textit{engineered} spin-orbit coupling due to a local, rotating magnetic field induced by a series of nanomagnets \cite{bib:Karmakar2011,bib:Klinovaja2013}. By adjusting either the period of the rotating field through the separation of the nanomagnets or again the length of the barriers, we can realize barriers with pure normal or spin-flip tunnelling. 

In summary, we conclude that the full Hamiltonian of our setup is given by $H=U_{C}+H_{SC}+H_{T}+H_{N}+H_{S}$.

\begin{figure}[!t] \centering
\includegraphics[width=\linewidth] {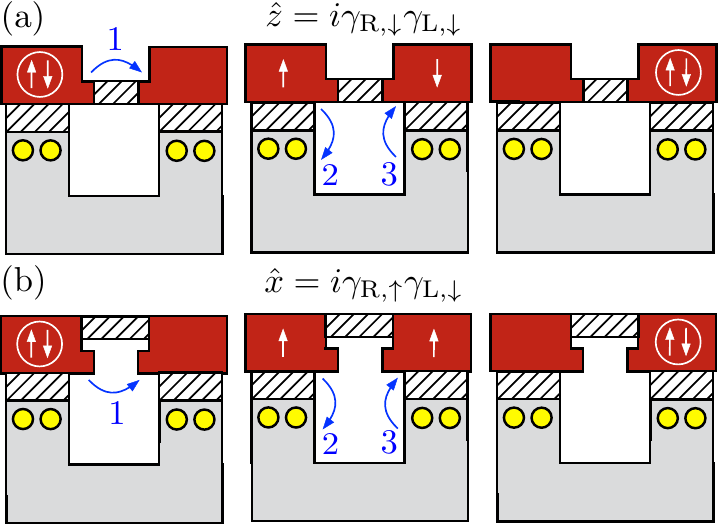}
\caption{(Color online)
(a) To third order in the tunnelling amplitudes, a Cooper pair moves between the two SC lead by splitting up between the normal tunnelling barrier and the TRI TSC island. The spin-flip tunnelling barrier is fully depleted by a local gate.  The resulting terms in the effective Hamiltonian are $\propto\hat{z}$. (b) Same as (a) but this time the Cooper pair splits up between the spin-flip tunnelling barrier and the TRI TSC island. The normal tunnelling barrier is fully depleted by a local gate. The resulting terms in the effective Hamiltonian are now $\propto\hat{x}$. 
}\label{fig:2}
\end{figure}

{\it Single-qubit Clifford gates.} 
In this section, we will implement single-qubit Clifford gates by ``Majorana tracking" \cite{bib:Litinski2017}.  This means that for a given circuit of single-qubit Clifford gates we record all Pauli operator redefinitions on a classical computer and use the quantum hardware only to perform suitable measurements of the $\hat{x}, \hat{y}, \hat{z}$-Pauli operators at the end of the computation.  

As a starting point, for measuring the $\hat{z}$-Pauli operator, we consider the situation when a local gate depletes the spin-flip tunnelling barrier between the two SC leads such that $\text{Im}\, t_{S}=0$ for $n_0$ even and $\text{Re}\, t_{S}=0$ for $n_{0}$ odd. 

In this case, to second order in the tunnelling amplitudes $t_{N}$, Cooper pairs tunnel between the SC leads only via the normal tunnelling barrier inducing a finite Josephson coupling $J_{N}\sim |t_{N}|^{2}/\Delta$. In particular, a Josephson coupling due to Cooper tunnelling between each SC lead and the TRI TSC is highly unfavorable as a result of the substantial island charging energy \cite{bib:Schrade2018_2}. We, hence, recognize that the island charging energy plays two significant roles in our setup: First, it suppresses quasiparticle poisoning due to single-electron tunnelling from the environment. Second, it also suppresses local mixing terms $\propto\hat{y}$ due to Cooper pair tunnelling between each SC lead and the TRI TSC island. Such local mixing terms are -- as noticed in the previous literature \cite{bib:Wolms2014} -- of importance for TRI TSCs with zero charging energy and, as we will see, can be utilized for measuring the $\hat{y}$-Pauli operator. 

As a next step, we note that to third order in the tunnelling amplitudes $t_{N},\lambda_{\text{L}},\lambda_{\text{R}}$ Cooper pair splitting sequences between the TRI TSC island and the normal tunnelling barrier induce additional Josephson couplings, $J_{z}$ for $n_{0}$ even and $J'_{z}$ for $n_{0}$ odd. We depict an example of such a Cooper pair splitting sequence in Fig.~\ref{fig:2}(a). In a first process, a Cooper pair on the left SC lead breaks up and one of the electrons tunnels via the normal tunnelling barrier to the right SC lead. This leaves the left SC lead in an excited state with one quasiparticle above the SC gap. In a second process, the quasiparticle on the left SC tunnels to the TRI TSC island and increments its charge by one unit. While the left SC returns to its ground state in this way, the TRI TSC island is now in an excited state with one excess charge. It, therefore, requires a third process to remove the extra charge from the TRI TSC by recombining it to a Cooper pair on the right SC lead. Critically, the tunnelling events via both the normal tunnelling barrier and the TRI TSC island conserve the electron spin. For that reason, the just described third-order sequences contribute terms $\propto\hat{z}=i\gamma_{\text{R},\da}\gamma_{\text{L},\da}=(-1)^{n_{0}}i\gamma_{\text{L},\ua}\gamma_{\text{R},\ua}$. 

As the last step, we  point out that to fourth order in the tunnelling amplitudes, Cooper pairs tunnel between the two SC leads only via the TRI TSC island yielding a Josephson coupling $J$ with a sign determined by the joint parity of all four MBSs on the TRI TSC island \cite{bib:Schrade2018}.

In the limit of weak tunnel couplings, $\pi\nu_{\ell}\lambda_{\ell}^{2}\ll\Delta,e^{2}/2C$ with $\nu_{\ell}$ the normal-state density of states per spin of the $\ell$-SC lead at the Fermi energy, we compute the amplitudes of all above-mentioned sequences perturbatively. Up to fourth order in the tunnelling amplitudes, we then summarize our results in an effective Hamiltonians acting on the ground states of the SC leads and the TRI TSC island. For $n_{0}$ even and $n_{0}$ odd, we find that,
\begin{equation}
\begin{split}
&H_{z,\text{even}}=-(J_{N}-J+\hat{z}\, J_{z})\cos\varphi, 
\\
&H_{z,\text{odd}}=-(J_{N}+J)\cos\varphi +  \hat{z} \, J'_{z}  \sin\varphi, \label{Hz}
\end{split}
\end{equation}
where the detailed microscopic forms of the Josephson couplings  $J$ and $J_{z}, J'_{z}$ are given in \cite{bib:Schrade2018} and \cite{bib:supplemental}, respectively. Here, it suffices to remark that $J_{z}\neq0$ $(J'_{z}\neq0)$ provided $\text{Im}\, t_{N}\neq0$ ($\text{Re} \, t_{N}\neq0$). Moreover, we point out that the both effective Hamiltonian exhibit TRS: For $H_{z,\text{even}}$ both $\hat{z}$ and $\cos\varphi$ are time-reversal even, while for $H_{z,\text{odd}}$ both $\hat{z}$ and $\sin\varphi$ are time-reversal odd. 

To measure the $z$-eigenvalue  of the $\hat{z}$-Pauli operator, we adopt a two-step protocol: (1) First, we separately measure the Josephson current through the normal tunnelling barrier and through the TRI TSC island to determine $J_{N}$ and $J$. (2) Second, we measure the Josephson current through the entire device. For $n_{0}$ even, the latter is given by $I=I_{c}\sin\varphi$ with the critical current $I_c =2e(J_{N}-J+z\, J_{z})/\hbar$ fixing the $z$-eigenvalue. 
For $n_0$ odd, the current phase relation is of the form $I=I_{c}\sin(\varphi+\varphi_0)$. This time it is not the critical current $I_{c}=2e\,\text{sgn}(J_{N}+J)\sqrt{(J_{N}+J)^{2}+(J'_z)^{2}}/\hbar$ but the anomalous phase shift $\varphi_{0}=z\,\arctan[J'_{z}/(J_{N}+J)]$ which fixes the $z$-eigenvalue. We note that the finite anomalous phase shift results from the $\hat{z}$-eigenstates breaking TRS when $n_{0}$ odd.

For the measurement of the $\hat{x}$-Pauli operator, we now shift our attention to a fully depleted normal tunnelling barrier such that $\text{Im}\, t_{N}=0$ for $n_0$ even and $\text{Re}\, t_{N}=0$ for $n_{0}$ odd. Similar to our earlier discussions, second order co-tunnelling events induce a Josephson coupling $J_{S}\sim|t_{S}|^{2}/\Delta$ as a result of Cooper pair tunnelling via the spin-flip tunnelling barrier whereas fourth order co-tunnelling events mediate a Josephson coupling $J$ via the TRI TSC island. However, a qualitative difference to the preceding considerations arises for the third order Cooper pair splitting sequences, see Fig.~\ref{fig:2}(b). These sequences now demand two spin-flips, one for an electron to move through the spin-flip tunnelling barrier and one for an electron to move through the TRI TSC island. As a consequence, the third-order sequences now contribute terms $\propto\hat{x}=i\gamma_{\text{R},\ua}\gamma_{\text{L},\da}=(-1)^{n_0}i\gamma_{\text{R},\da}\gamma_{\text{L},\ua}$. More explicitly, up to fourth order in the weak tunnel couplings, the effective Hamiltonians for $n_0$ even and $n_0$ odd are given by, 
\begin{equation}
\begin{split}
&H_{x,\text{even}}=-(J_{N}-J+\hat{x}\, J_{x})\cos\varphi, 
\\
&H_{x,\text{odd}}=-(J_{N}+J)\cos\varphi +  \hat{x} \, J'_{z}  \sin\varphi. \label{Hx}
\end{split}
\end{equation}
Here, $J_{x}, J'_{x}$ denote the Josephson couplings due to the Cooper pair splitting processes.  For their microscopic form, see \cite{bib:supplemental}. Here, we only remark that $J_{x}\neq 0$ ($J'_{x}\neq 0$) granted that $\text{Im}\,t_{S}\neq0$ ($\text{Re}\,t_{S}\neq0$). We further point out that both effective Hamiltonian exhibit TRS: For $H_{x,\text{even}}$ both $\hat{x}$ and $\cos\varphi$ are time-reversal even, while for $H_{x,\text{odd}}$ both $\hat{x}$ and $\sin\varphi$ are time-reversal odd. To measure the $\hat{x}$-Pauli operator, we readily see that the effective Hamiltonians are of the same form as those in Eq.~\eqref{Hz}. Consequently, our previously introduced measurement protocol for the  $\hat{z}$-Pauli operator immediately carries over to the measurement of the $\hat{x}$-Pauli operator. 

At this point, it is worth mentioning that a potential error source for the $\hat{x}$, $\hat{z}$-measurements occurs when both
$\text{Im}\, t_{N}\neq0$, $\text{Im}\, t_{S}\neq0$ for $n_0$ even or $\text{Re}\, t_{N}\neq0$, $\text{Re}\, t_{S}\neq0$ for $n_{0}$ odd. In practice, this happens either when one of the tunnelling barriers is not fully depleted, or the barrier lengths $d,d'$ are not appropriately adjusted to the spin-orbit length $\lambda_{\text{SO}}$. Fortunately, this constitutes a static hardware error which can be addressed prior to all experiments. In particular, the error can be made controllably small with a careful design of a \textit{conventional} Josephson junction.

In the last part of this section, we address $\hat{y}$-measurements. These require the tuning the charging energy of the TRI TSC to zero which is attainable -- on demand -- by coupling the TRI TSC island to a bulk SC through a gate-tunable valve \cite{bib:Aasen2016}. Critically, even at zero charging energy the value of the joint fermion parity in Eq.~\eqref{TotalParity} remains protected as a result of the lead SC gap. However, unlike in the case of a substantial charging energy, Cooper pairs can now tunnel in a second order process between each SC lead and the TRI TSC island inducing a Josephson coupling  $\propto\hat{y}$ \cite{bib:Chung2013}. Consequently, the resulting Josephson current provides a means for measuring the $\hat{y}$ eigenvalue. The details of this measurement scheme are discussed in \cite{bib:Chung2013}. 

{\it Universal quantum computation.}
So far, we have discussed the implementation of single-qubit Clifford gates. However, for universal quantum computation, we need to supplement the single-qubit Clifford gates by a single-qubit $\pi/8$-gate and a two-qubit entangling gate \cite{bib:Brylinski2001}. Clearly, by pulsing the Josephson couplings in the effective Hamiltonians of Eq.~\eqref{Hz} and \eqref{Hx} we can perform arbitrary rotations on the MKQ Bloch sphere and, therefore, in particular, a $\pi/8$-gate. For this procedure, phase-independent contributions -- which were irrelevant for the Josephson current -- should now be included in the effective Hamiltonians, see \cite{bib:supplemental}.

\begin{figure}[!t] \centering
\includegraphics[width=0.8\linewidth] {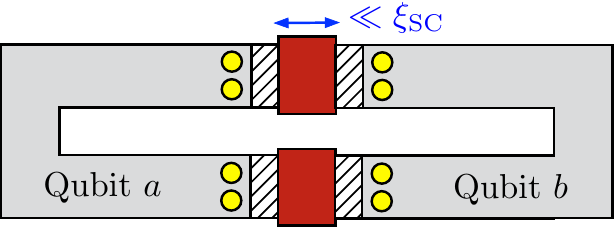}
\caption{(Color online)
Setup of two MKQs $a,b$ coupled to two SC leads. The width of the leads is much smaller than their SC coherence length $\xi_{\text{SC}}$, thereby, permitting Cooper pair splitting between the two MKQs. The resulting anisotropic Heisenberg interaction between the two MKQs is used to construct a two-qubit entangling gate. 
}\label{fig:3}
\end{figure}

For a two-qubit entangling gate, we consider the setup of Fig.~\ref{fig:3} which comprises two SC leads addressing two MKQs $a,b$. Here, a local gate fully depletes both the  normal and the spin-flip tunnelling barrier,  $t_{N}=t_{S}=0$. Provided that the width of the SC leads is much smaller than the SC coherence length $\xi_{\text{SC}}$, a Cooper pair can now split up between the two TRI TSC islands and generate entanglement between the MKQs. For symmetric couplings and a ground state charge $n_0$ for both TRI TSC islands, we have computed the amplitudes of these processes in the weak coupling limit. An effective anisotropic Heisenberg interaction summarizes the results, 
\begin{equation}
\begin{split}
\label{HEff_Two_Qubit}
H_{ab} &=J_{y}\hat{y}_{a}\hat{y}_{b}\\
&+[J_{xz}+(-1)^{n_{0}+1}J'_{xz}\cos\varphi] (\hat{x}_{a}\hat{x}_{b}+\hat{z}_{a}\hat{z}_{b}).
\end{split}
\end{equation}
For the microscopic form of the coupling constants $J_{xz},J'_{xz},J_{y}$, see \cite{bib:supplemental}. We note that Heisenberg interaction can be made isotropic by choosing the SC phase difference such that $\tilde{J}\equiv J_{y}=J_{xz}+(-1)^{n_{0}+1}J'_{xz}\cos\varphi$. Pulsing the couplings for a duration $\tau$ defined by  
$
\int^{\tau}_{0} \tilde{J}(t')\ \mathrm{d}t'= \pi/2 \ (\text{mod}\ \pi)
$
then implements a $\sqrt{\text{SWAP}}$-gate via the unitary time evolution operator. The latter, in combination with single-qubit gates, is sufficient for universal quantum computing \cite{bib:Loss1998}. 

{\it Conclusions.}
In this work, we have introduced a qubit formed by the degenerate ground states of a Coulomb-blockaded TRI TSC island with spatially well separated MKPs: the ``Majorana Kramers Qubit". By coupling a single MKQ to SC leads, we have shown that in principle single-qubit Clifford gates can be realized though qubit measurements. Furthermore, we argued that a $\pi/8$-gate as well as a two-MKQ entangling gate can be realized by pulsing of Josephson couplings. 
Besides providing the conceptual insight that strong magnetic fields are not required for Majorana-based quantum computing, we hope that the MKQ will also provide an alternative route towards a robust quantum computer.

{\it Acknowledgments.}
We would like to thank Jagadeesh S. Moodera for helpful discussions.
C.S. was supported by the Swiss SNF under Project 174980. L.F. and C.S. were supported by DOE Office of Basic Energy Sciences, Division of Materials Sciences and Engineering under Award $\text{DE-SC0010526}$.

\begin{widetext}

\newpage

\onecolumngrid

\bigskip 

\begin{center}
\large{\bf Supplemental Material to `Quantum Computing with Majorana Kramers Pairs' \\}
\end{center}
\begin{center}
Constantin Schrade and Liang Fu
\\
{\it Department of Physics, Massachusetts Institute of Technology, 77 Massachusetts Ave., Cambridge, MA 02139}
\end{center}

In the Supplemental Material, we provide more details on the derivation of the effective Hamiltonians required for the implementation of the single- and two-qubit quantum gates.

\section{Effective Hamiltonian for the single-qubit gates}
In this first section of the Supplemental Material,  we derive the effective Hamiltonians for the single-qubit gates as given in Eq.~(9) and (10) of the main text. For simplicity, we adopt two assumptions: First, we assume that the lead SC gaps are of equal magnitude, $\Delta\equiv\Delta_{\ell}$. Second, we assume that the gate charge of the TRI TSC island $Q/e$ is tuned to an integer value $n_{0}$. In this, so-called, Coulomb-valley regime the ground state of the TRI TSC island consists of $n_{0}$ units of charge. At the same time, adding/removing a single unit of charge from the TRI TSC island comes at an equal energy cost of $U\equiv e^{2}/2C$.
\\

Initially, we derive the effective Hamiltonian given in Eq.~(9) of the main text that is used for the measurement of the $\hat{z}$-Pauli operator. Our focus are the contributions to the effective Hamiltonian which are most important for the measurement protocol and which occur at third-order in the tunnelling amplitudes. They are given by, 
\begin{equation}
\begin{split}
H^{(3)}_{z}&=
- P_{n_0} H_{T,z} \left(\left[H_{0}+U_{C}\right]^{-1}\left[1-P_{n_0}\right]H_{T,z}\right)^{2}P_{n_0}.
\label{H3z}
\end{split}
\end{equation}
Here, $H_{0}=U_{C}+H_{\text{SC}}$ denotes the Hamiltonian of the uncoupled system and $H_{T,z}=H_{T}+H_{N}$ denotes the total tunnelling Hamiltonian. For the moment, we have set $t_{S}=0$. However, after having derived both effective Hamiltonians given in Eqs.~(9) and (10) of the main text, we will see that it is sufficient to require $\text{Im}(t_{S})=0$ for $n_0$ even and $\text{Re}(t_{S})=0$ for $n_{0}$ odd. Finally, we note that $P_{n_0}=\Pi_{n_0}\Pi_{\text{BCS}}$ where $\Pi_{n_0}$ is a projector on the ground state of the TRI TSC island with $n_0$ units of charge and $\Pi_{\text{BCS}}$ is a projector on the BCS (Bardeen-Cooper-Schrieffer) ground states of the SC leads. 

As the first step in our derivation, we interpret Eq.~\eqref{H3z} as the weighted sum of all three-step sequences of intermediate states that map the ground state manifold of $H_{0}$ onto itself. For the moment, we will only discuss the three-step sequences which comprise the transport of a Cooper pair from the left to the right SC lead or vice versa, see Fig.~\ref{fig:1_SM}. Two example sequences of the type shown in Fig.~\ref{fig:1_SM}(a) are, 
\begin{equation}
\begin{split}
&\quad\
P_{n_{0}}
(
\lambda
c^{\dag}_{\text{R},\bk \ua}
\gamma_{\text{R},\ua}
e^{-i\phi/2}
)
(
\lambda
\gamma_{\text{L},\ua}
c_{\text{L},\bk \ua}
e^{i\phi/2}
)
(
t^{*}_{N}c^{\dag}_{\text{R},-\bk \da}c_{\text{L},-\bk \da} 
)
P_{n_{0}}
\\
&=
t^{*}_{N}\lambda^{2}
P_{n_{0}}
(
c^{\dag}_{\text{R},\bk \ua}
\gamma_{\text{R},\ua}
\gamma_{\text{L},\ua}
c_{\text{L},\bk \ua}
c^{\dag}_{\text{R},-\bk \da}c_{\text{L},-\bk \da} 
)
P_{n_{0}}
\\
&=
(-1)^{n_{0}+1}
t^{*}_{N}\lambda^{2}
(
\Pi_{n_0}
\gamma_{\text{R},\da}
\gamma_{\text{L},\da}
\Pi_{n_0}
)
(
\Pi_{\text{BCS}}
c^{\dag}_{\text{R},\bk \ua}
c_{\text{L},\bk \ua}
c^{\dag}_{\text{R},-\bk \da}c_{\text{L},-\bk \da} 
\Pi_{\text{BCS}})
\\
&=
i(-1)^{n_{0}}
t^{*}_{N}\lambda^{2}
(
\Pi_{n_0}
\hat{z}
\Pi_{n_0}
)
(
\Pi_{\text{BCS}}
c^{\dag}_{\text{R},\bk \ua}
c_{\text{L},\bk \ua}
c^{\dag}_{\text{R},-\bk \da}c_{\text{L},-\bk \da} 
\Pi_{\text{BCS}})
\\
&=
i(-1)^{n_{0}+1}
e^{i(\varphi_{\text{L}}-\varphi_{\text{R}})}
t^{*}_{N}\lambda^{2}
(u_{\bk}v_{\bk})^{2}
(
\Pi_{n_0}
\hat{z}
\Pi_{n_0}
)
(
\Pi_{\text{BCS}}
\gamma_{\text{R},-\bk\da}
\gamma_{\text{L},\bk\ua}
\gamma^{\dag}_{\text{R},-\bk\da}
\gamma^{\dag}_{\text{L},\bk\ua}
\Pi_{\text{BCS}})
\\
&=
i
(-1)^{n_{0}}
e^{i(\varphi_{\text{L}}-\varphi_{\text{R}})}
t^{*}_{N}\lambda^{2}
(u_{\bk}v_{\bk})^{2}
(
\Pi_{n_0}
\hat{z}
\Pi_{n_0}
)
\Pi_{\text{BCS}}
\\
\\
&\quad\
P_{n_{0}}
(
\lambda
c^{\dag}_{\text{R},-\bk \da}
\gamma_{\text{R},\da}
e^{-i\phi/2}
)
(
\lambda
\gamma_{\text{L},\da}
c_{\text{L},-\bk \da}
e^{i\phi/2}
)
(
t_{N}c^{\dag}_{\text{R},\bk \ua}c_{\text{L},\bk \ua} 
)
P_{n_{0}}
\\
&=
t_{N}\lambda^{2}
P_{n_{0}}
(
c^{\dag}_{\text{R},-\bk \da}
\gamma_{\text{R},\da}
\gamma_{\text{L},\da}
c_{\text{L},-\bk \da}
c^{\dag}_{\text{R},\bk \ua}c_{\text{L},\bk \ua} 
)
P_{n_{0}}
\\
&=
t_{N}\lambda^{2}
(
\Pi_{n_0}
\gamma_{\text{R},\da}
\gamma_{\text{L},\da}
\Pi_{n_0}
)
(
\Pi_{\text{BCS}}
c^{\dag}_{\text{R},-\bk \da}
c_{\text{L},-\bk \da}
c^{\dag}_{\text{R},\bk \ua}
c_{\text{L},\bk \ua}
\Pi_{\text{BCS}} 
)
\\
&=
-i
t_{N}\lambda^{2}
(
\Pi_{n_0}
\hat{z}
\Pi_{n_0}
)
(
\Pi_{\text{BCS}}
c^{\dag}_{\text{R},-\bk \da}
c_{\text{L},-\bk \da}
c^{\dag}_{\text{R},\bk \ua}
c_{\text{L},\bk \ua}
\Pi_{\text{BCS}} 
)
\\
&=
i
t_{N}\lambda^{2}
e^{i(\varphi_{\text{L}}-\varphi_{\text{R}})}
(u_{\bk}v_{\bk})^{2}
(
\Pi_{n_0}
\hat{z}
\Pi_{n_0}
)
(
\Pi_{\text{BCS}}
\gamma_{\text{R},\bk\ua}
\gamma_{\text{L},-\bk\da}
\gamma^{\dag}_{\text{R},\bk\ua}
\gamma^{\dag}_{\text{L},-\bk\da}
\Pi_{\text{BCS}} 
)
\\
&=
-it_{N}\lambda^{2}
e^{i(\varphi_{\text{L}}-\varphi_{\text{R}})}
(u_{\bk}v_{\bk})^{2}
(
\Pi_{n_0}
\hat{z}
\Pi_{n_0}
)
\Pi_{\text{BCS}} 
\end{split}
\end{equation}
We remark that in the fifth line of both calculations, we have rewritten the electron operators of SC leads in terms of Bogoliubov quasiparticles, $c_{\ell,\bk \ua}=e^{i\varphi_{\ell}/2}(u_{\bk}\gamma_{\ell,\bk \ua}+v_{\bk}\gamma^{\dag}_{\ell,-\bk \da})$ and $c_{\ell,-\bk \da}=e^{i\varphi_{\ell}/2}(u_{\bk}\gamma_{\ell,-\bk \da}-v_{\bk}\gamma^{\dag}_{\ell,\bk \ua})$. 
Adding these two sequences as well as their hermitian-conjugated counterparts, multiplying by the energy denominator $-1/(E_{\bk}+U)(2E_{\bk})$ and carrying out the summation over all momenta, yields the contribution,
\begin{equation}
\begin{split}
&-\hat{z} J^{\text{(a)}}_{z} \cos\varphi \quad \text{with} \quad J^{\text{(a)}}_{z}=4\lambda^{2}\,\text{Im}(t_{N})
\sum_{\bk}\frac{(u_{\bk}v_{\bk})^{2}}{(E_{\bk}+U)E_{\bk}} \quad \text{for} \ n_{0} \ \text{even}, 
\\
&\quad\  \hat{z} J'^{\text{(a)}}_{z} \sin\varphi \quad  \text{with} \quad J'^{\text{(a)}}_{z}=-4\lambda^{2}\,\text{Re}(t_{N})
\sum_{\bk}\frac{(u_{\bk}v_{\bk})^{2}}{(E_{\bk}+U)E_{\bk}} \quad \text{for} \ n_{0} \ \text{odd}.
\end{split}
\end{equation}
Here, for notational brevity, we have dropped the projectors on the ground state manifold of $H_{0}$. 
As the second step, we evaluate the Cooper pair transport sequences corresponding to Figs.~\ref{fig:1_SM}(b) to (f) in a similar way. This gives, 
\begin{equation}
\begin{split}
&-\hat{z} J_{z} \cos\varphi \quad \text{with} \quad J_{z}=J^{\text{(a)}}_{z}+J^{\text{(b)}}_{z}+J^{\text{(c)}}_{z}+J^{\text{(d)}}_{z}+J^{\text{(e)}}_{z}+J^{\text{(f)}}_{z} \quad \text{for} \ n_{0} \ \text{even}, 
\\
&\hspace{12pt}\hat{z} J'_{z} \sin\varphi \quad \text{with} \quad J'_{z}=J'^{\text{(a)}}_{z}+J'^{\text{(b)}}_{z}+J'^{\text{(c)}}_{z}+J'^{\text{(d)}}_{z}+J'^{\text{(e)}}_{z}+J'^{\text{(f)}}_{z} \quad \text{for} \ n_{0} \ \text{odd},
\end{split}
\end{equation}
where we have introduced the coupling constants,
\begin{equation}
\begin{split}
&J^{\text{(a)}}_{z}=J^{\text{(b)}}_{z}=J^{\text{(e)}}_{z}=J^{\text{(f)}}_{z}
\quad , \quad 
J^{\text{(c)}}_{z}=J^{\text{(d)}}_{z}=8\lambda^{2}\,\text{Im}(t_{N})
\sum_{\bk}\left(\frac{u_{\bk}v_{\bk}}{E_{\bk}+U}\right)^{2},
\\
&J'^{\text{(a)}}_{z}=J'^{\text{(b)}}_{z}=J'^{\text{(e)}}_{z}=J'^{\text{(f)}}_{z}
\quad , \quad 
J'^{\text{(c)}}_{z}=J'^{\text{(d)}}_{z}=-8\lambda^{2}\,\text{Re}(t_{N})
\sum_{\bk}\left(\frac{u_{\bk}v_{\bk}}{E_{\bk}+U}\right)^{2}.
\end{split}
\end{equation}
Adding the second order contribution, which corresponds to a conventional Josephson effect through the normal tunnelling barrier, as well as the fourth order contribution, which corresponds to a parity-controlled $2\pi$-Josephson effect through the TRI TSC island \cite{bib:Schrade2018_SM}, we arrive at the effective Hamiltonian given in Eq.~(9) of the main text,
\begin{equation}
\begin{split}
&H_{z,\text{even}}=-(J_{N}-J+\hat{z}\, J_{z})\cos\varphi \quad, \quad H_{z,\text{odd}}=-(J_{N}+J)\cos\varphi +  \hat{z} \, J'_{z}  \sin\varphi. 
\end{split}
\end{equation}
Up to this point, we have only considered Cooper pair transport sequences which induce terms proportional to the SC phase difference in the effective Hamiltonian. Terms that are independent of the SC phase difference do not modify the Josephson current and, for that reason, have been omitted for the $\hat{z}$-measurement protocol. However, terms which are independent of the SC phase difference are of relevance when pulsing the tunnel couplings to obtain a $\pi/8$-gate. For that reason, we now provide a derivation of these contributions. First, we again examine two example sequence of the type shown in Fig.~\ref{fig:1_SM}(a), 
\begin{equation}
\begin{split}
&\quad\
P_{n_{0}}
(
\lambda
\gamma_{\text{R},\da}
c_{\text{R},-\bk \da}
e^{i\phi/2}
)
(
\lambda
c^{\dag}_{\text{L},-\bk \da}
\gamma_{\text{L},\da}
e^{-i\phi/2}
)
(
t^{*}_{N}c^{\dag}_{\text{R},-\bk \da}c_{\text{L},-\bk \da} 
)
P_{n_{0}}
\\
&
=
t^{*}_{N}\lambda^{2}
P_{n_{0}}
(
\gamma_{\text{R},\da}
c_{\text{R},-\bk \da}
c^{\dag}_{\text{L},-\bk \da}
\gamma_{\text{L},\da}
c^{\dag}_{\text{R},-\bk \da}
c_{\text{L},-\bk \da} 
)
P_{n_{0}}
\\
&=
t^{*}_{N}\lambda^{2}
(\lambda
\Pi_{n_0}
\gamma_{\text{R},\da}
\gamma_{\text{L},\da}
\Pi_{n_0}
)
(
\Pi_{\text{BCS}}
c_{\text{R},-\bk \da}
c^{\dag}_{\text{L},-\bk \da}
c^{\dag}_{\text{R},-\bk \da}
c_{\text{L},-\bk \da} 
\Pi_{\text{BCS}})
\\
&=
-i
t^{*}_{N}\lambda^{2}
(
\Pi_{n_0}
\hat{z}
\Pi_{n_0}
)
(
\Pi_{\text{BCS}}
c_{\text{R},-\bk \da}
c^{\dag}_{\text{L},-\bk \da}
c^{\dag}_{\text{R},-\bk \da}
c_{\text{L},-\bk \da} 
\Pi_{\text{BCS}})
\\
&=
-i
t^{*}_{N}\lambda^{2}
(u_{\bk}v_{\bk})^{2}
(
\Pi_{n_0}
\hat{z}
\Pi_{n_0}
)
(
\Pi_{\text{BCS}}
\gamma_{\text{R},-\bk\da}
\gamma_{\text{L},\bk\ua}
\gamma^{\dag}_{\text{R},-\bk\da}
\gamma^{\dag}_{\text{L},\bk\ua}
\Pi_{\text{BCS}})
\\
&=
i
t^{*}_{N}\lambda^{2}
(u_{\bk}v_{\bk})^{2}
(
\Pi_{n_0}
\hat{z}
\Pi_{n_0}
)
\Pi_{\text{BCS}},
\\
\\
&\quad\
P_{n_{0}}
(
\lambda
\gamma_{\text{R},\ua}
c_{\text{R},\bk \ua}
e^{i\phi/2}
)
(
\lambda
c^{\dag}_{\text{L},\bk \ua}
\gamma_{\text{L},\ua}
e^{-i\phi/2}
)
(
t_{N}c^{\dag}_{\text{R},\bk \ua}c_{\text{L},\bk \ua} 
)
P_{n_{0}}
\\
&=
t_{N}\lambda^{2}
P_{n_{0}}
(
\gamma_{\text{R},\ua}
c_{\text{R},\bk \ua}
c^{\dag}_{\text{L},\bk \ua}
\gamma_{\text{L},\ua}
c^{\dag}_{\text{R},\bk \ua}
c_{\text{L},\bk \ua} 
)
P_{n_{0}}
\\
&=
(-1)^{n_{0}+1}
t_{N}\lambda^{2}
(
\Pi_{n_0}
\gamma_{\text{R},\da}
\gamma_{\text{L},\da}
\Pi_{n_0}
)
(
\Pi_{\text{BCS}}
c_{\text{R},\bk \ua}
c^{\dag}_{\text{L},\bk \ua}
c^{\dag}_{\text{R},\bk \ua}
c_{\text{L},\bk \ua} 
\Pi_{\text{BCS}} 
)
\\
&=
i(-1)^{n_{0}}
t_{N}\lambda^{2}
(
\Pi_{n_0}
\hat{z}
\Pi_{n_0}
)
(
\Pi_{\text{BCS}}
c_{\text{R},\bk \ua}
c^{\dag}_{\text{L},\bk \ua}
c^{\dag}_{\text{R},\bk \ua}
c_{\text{L},\bk \ua} 
\Pi_{\text{BCS}} 
)
\\
&=
i(-1)^{n_{0}}
t_{N}\lambda^{2}
(u_{\bk}v_{\bk})^{2}
(
\Pi_{n_0}
\hat{z}
\Pi_{n_0}
)
(
\Pi_{\text{BCS}}
\gamma_{\text{R},\bk\ua}
\gamma_{\text{L},-\bk\da}
\gamma^{\dag}_{\text{R},\bk\ua}
\gamma^{\dag}_{\text{L},-\bk\da}
\Pi_{\text{BCS}} 
)
\\
&=
i(-1)^{n_{0}+1}
t_{N}\lambda^{2}
(u_{\bk}v_{\bk})^{2}
(
\Pi_{n_0}
\hat{z}
\Pi_{n_0}
)
\Pi_{\text{BCS}} 
\end{split}
\end{equation}
\begin{figure}[!t] \centering
\includegraphics[width=\linewidth] {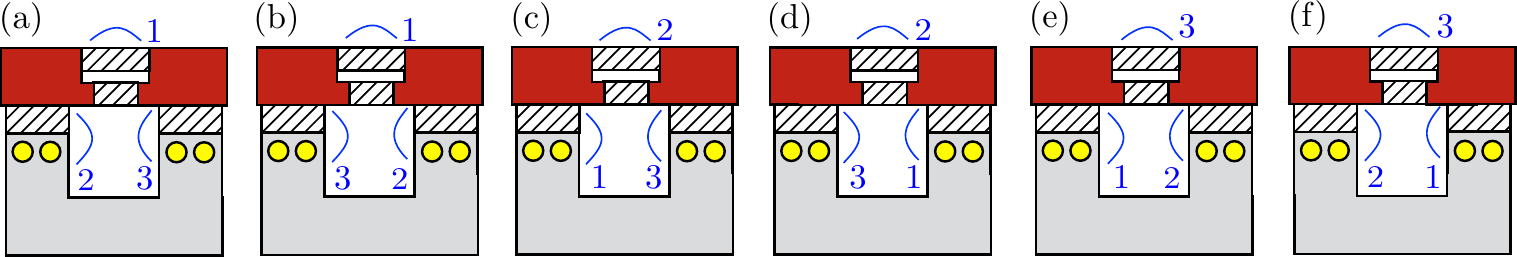}
\caption{(Color online)
Third-order sequences of intermediate states (up to hermitian-conjugation) relevant for the single-qubit gates. 
The enumerated blue lines label the tunnel couplings which are turned on for a given intermediate step within a third-order sequence. 
}\label{fig:1_SM}
\end{figure}
When combining the two sequences with their hermitian-conjugated counterparts, their contributions cancel each other for $n_{0}$ odd but yield a finite contribution for $n_{0}$ even. Incorporating all possible contributions, as shown in Figs.~\ref{fig:1_SM}(a) to (f), we find that for $n_{0}$ even the effective Hamiltonian changes to, 
\begin{equation}
\begin{split}
&H_{z,\text{even}}\rightarrow\hat{z}\tilde{J}_{z}-(J_{N}-J+\hat{z}\, J_{z})\cos\varphi \quad \text{with} \quad  
\tilde{J}_{z}=\tilde{J}_{z,a}+\tilde{J}_{z,b}+\tilde{J}_{z,c}+\tilde{J}_{z,d}+\tilde{J}_{z,e}+\tilde{J}_{z,f}. 
\end{split}
\end{equation}
Here, we have defined additional coupling constants,
\begin{equation}
\begin{split}
&\tilde{J}_{z,a}=\tilde{J}_{z,b}=\tilde{J}_{z,e}=\tilde{J}_{z,f}=-4\lambda^{2}\,\text{Im}(t_{N})
\sum_{\bk}\frac{(u_{\bk}v_{\bk})^{2}}{(E_{\bk}+U)E_{\bk}}
\quad , \quad 
\tilde{J}_{z,c}=\tilde{J}_{z,d}=8\lambda^{2}\,\text{Im}(t_{N})
\sum_{\bk}\left(\frac{u_{\bk}v_{\bk}}{E_{\bk}+U}\right)^{2}.
\end{split}
\end{equation}
We again remark that the effective Hamiltonian for $n_{0}$ odd remains unchanged as a result of the cancellations discussed above.
\\

We now proceed by deriving the effective Hamiltonian given in Eq.~(10) of the main text that is used for measuring the $\hat{x}$-Pauli operator. The all-important third-order contributions to the effective Hamiltonian are given by,
\begin{equation}
\begin{split}
H^{(3)}_{x}&=
- P H_{T,x} \left(\left[H_{0}+U_{C}\right]^{-1}\left[1-P\right]H_{T,x}\right)^{2}P, 
\label{H3x}
\end{split}
\end{equation}
where $H_{T,x}=H_{T}+H_{S}$ denotes the total tunnelling Hamiltonian. For now, we have again set $t_{N}=0$. However, it will become clear after this derivation that it is sufficient to require $\text{Im}(t_{N})=0$ for $n_0$ even and $\text{Re}(t_{N})=0$ for $n_{0}$ odd. To begin, we consider two example sequences of the type shown in Fig.~\ref{fig:1_SM}(a),
\begin{equation}
\begin{split}
&\quad\
P_{n_{0}}
(
\lambda
c^{\dag}_{\text{R},\bk \da}
\gamma_{\text{R},\da}
e^{-i\phi/2}
)
(
\lambda
\gamma_{\text{L},\ua}
c_{\text{L},\bk \ua}
e^{-i\phi/2}
)
(
t_{S}c^{\dag}_{\text{R},-\bk \ua}c_{\text{L},-\bk \da} 
)
P_{n_{0}}
\\
&=
t_{S}\lambda^{2}
P_{n_{0}}
(
c^{\dag}_{\text{R},\bk \da}
\gamma_{\text{R},\da}
\gamma_{\text{L},\ua}
c_{\text{L},\bk \ua}
c^{\dag}_{\text{R},-\bk \ua}c_{\text{L},-\bk \da} 
)
P_{n_{0}}
\\
&=
t_{S}\lambda^{2}
(
\Pi_{n_0}
\gamma_{\text{R},\da}
\gamma_{\text{L},\ua}
\Pi_{n_0}
)
(
\Pi_{\text{BCS}}
c^{\dag}_{\text{R},\bk \da}
c_{\text{L},\bk \ua}
c^{\dag}_{\text{R},-\bk \ua}
c_{\text{L},-\bk \da} 
\Pi_{\text{BCS}}
)
\\
&=
i(-1)^{n_{0}+1}t_{S}\lambda^{2}
(
\Pi_{n_0}
\hat{x}
\Pi_{n_0}
)
(
\Pi_{\text{BCS}}
c^{\dag}_{\text{R},\bk \da}
c_{\text{L},\bk \ua}
c^{\dag}_{\text{R},-\bk \ua}
c_{\text{L},-\bk \da} 
\Pi_{\text{BCS}}
)
\\
&=
i(-1)^{n_{0}+1}
e^{i(\varphi_{\text{L}}-\varphi_{\text{R}})}
t_{S}\lambda^{2}
(u_{\bk}v_{\bk})^{2}
(
\Pi_{n_0}
\hat{x}
\Pi_{n_0}
)
(
\Pi_{\text{BCS}}
\gamma_{\text{R},-\bk\ua}
\gamma_{\text{L},\bk\ua}
\gamma^{\dag}_{\text{R},-\bk\ua}
\gamma^{\dag}_{\text{L},\bk\ua}
\Pi_{\text{BCS}})
\\
&=
i(-1)^{n_{0}}
e^{i(\varphi_{\text{L}}-\varphi_{\text{R}})}
t_{S}\lambda^{2}
(u_{\bk}v_{\bk})^{2}
(
\Pi_{n_0}
\hat{x}
\Pi_{n_0}
)
\Pi_{\text{BCS}},
\\
\\
&\quad\
P_{n_{0}}
(
\lambda
c^{\dag}_{\text{R},-\bk \ua}
\gamma_{\text{R},\ua}
e^{-i\phi/2}
)
(
\lambda
\gamma_{\text{L},\da}
c_{\text{L},-\bk \da}
e^{i\phi/2}
)
(
-
t^{*}_{S}c^{\dag}_{\text{R},\bk \da}c_{\text{L},\bk \ua} 
)
P_{n_{0}}
\\
&=
-
t^{*}_{S}\lambda^{2}
P_{n_{0}}
(
c^{\dag}_{\text{R},-\bk \ua}
\gamma_{\text{R},\ua}
\gamma_{\text{L},\da}
c_{\text{L},-\bk \da}
c^{\dag}_{\text{R},\bk \da}
c_{\text{L},\bk \ua} 
)
P_{n_{0}}
\\
&=
-
t^{*}_{S}\lambda^{2}
(
\Pi_{n_0}
\gamma_{\text{R},\ua}
\gamma_{\text{L},\da}
\Pi_{n_0}
)
(
\Pi_{\text{BCS}}
c^{\dag}_{\text{R},-\bk \ua}
c_{\text{L},-\bk \da}
c^{\dag}_{\text{R},\bk \da}
c_{\text{L},\bk \ua} 
\Pi_{\text{BCS}}
)
\\
&=
i
t^{*}_{S}\lambda^{2}
(
\Pi_{n_0}
\hat{x}
\Pi_{n_0}
)
(
\Pi_{\text{BCS}}
c^{\dag}_{\text{R},-\bk \ua}
c_{\text{L},-\bk \da}
c^{\dag}_{\text{R},\bk \da}
c_{\text{L},\bk \ua} 
\Pi_{\text{BCS}}
)
\\
&=
i
t^{*}_{S}\lambda^{2}
(
\Pi_{n_0}
\hat{x}
\Pi_{n_0}
)
(
\Pi_{\text{BCS}}
c^{\dag}_{\text{R},\bk \da}
c_{\text{L},\bk \ua}
c^{\dag}_{\text{R},-\bk \ua}
c_{\text{L},-\bk \da} 
\Pi_{\text{BCS}}
)
\\
&=
ie^{i(\varphi_{\text{L}}-\varphi_{\text{R}})}
t^{*}_{S}\lambda^{2}
(u_{\bk}v_{\bk})^{2}
(
\Pi_{n_0}
\hat{x}
\Pi_{n_0}
)
(
\Pi_{\text{BCS}}
\gamma_{\text{R},-\bk\ua}
\gamma_{\text{L},\bk\ua}
\gamma^{\dag}_{\text{R},-\bk\ua}
\gamma^{\dag}_{\text{L},\bk\ua}
\Pi_{\text{BCS}}
)
\\
&=
-
ie^{i(\varphi_{\text{L}}-\varphi_{\text{R}})}
t^{*}_{S}\lambda^{2}
(u_{\bk}v_{\bk})^{2}
(
\Pi_{n_0}
\hat{x}
\Pi_{n_0}
)
\Pi_{\text{BCS}}
\end{split}
\end{equation}
Here, we have made the inessential assumption that the normal state dispersion of the SC obeys $\xi_{\bk}=\xi_{-\bk}$ 
such that $u_{\bk}=u_{-\bk}$, $v_{\bk}=v_{-\bk}$ and $E_{\bk}=E_{-\bk}$. Adding the two sequencesas well as their hermitian-conjugated counterparts, multiplying by the energy denominator $-1/(E_{\bk}+U)(2E_{\bk})$ and carrying out the summation over all momenta, gives the contribution,
\begin{equation}
\begin{split}
&-\hat{x} J^{(\text{a})}_{x} \cos\varphi \quad \text{with} \quad J^{(\text{a})}_{x}=-4\lambda^{2}\,\text{Im}(t_{S})
\sum_{\bk}\frac{(u_{\bk}v_{\bk})^{2}}{(E_{\bk}+U)E_{\bk}} \quad \text{for} \ n_{0} \ \text{even}, 
\\
&\quad\  \hat{x} J'^{(\text{a})}_{x} \sin\varphi \quad  \text{with} \quad J'^{(\text{a})}_{x} =-4\lambda^{2}\,\text{Re}(t_{S})
\sum_{\bk}\frac{(u_{\bk}v_{\bk})^{2}}{(E_{\bk}+U)E_{\bk}} \quad \text{for} \ n_{0} \ \text{odd}.
\end{split}
\end{equation}
For notational brevity, we have again omitted the projectors on the ground state manifold of $H_{0}$. 
In a similar way, we also evaluate the Cooper pair transport sequences given in Fig.~\ref{fig:1_SM}(b) to (f). This yields,
\begin{equation}
\begin{split}
&-\hat{x} J_{x} \cos\varphi \quad \text{with} \quad J_{x}=J^{\text{(a)}}_{x}+J^{\text{(b)}}_{x}+J^{\text{(c)}}_{x}+J^{\text{(d)}}_{x}+J^{\text{(e)}}_{x}+J^{\text{(f)}}_{x} \quad \text{for} \ n_{0} \ \text{even}, 
\\
&\hspace{12pt}\hat{z} J'_{x} \sin\varphi \quad \text{with} \quad J'_{x}=J'^{\text{(a)}}_{x}+J'^{\text{(b)}}_{x}+J'^{\text{(c)}}_{x}+J'^{\text{(d)}}_{x}+J'^{\text{(e)}}_{x}+J'^{\text{(f)}}_{x} \quad \text{for} \ n_{0} \ \text{odd}.
\end{split}
\end{equation}
Here, we have defined the coupling constants,
\begin{equation}
\begin{split}
&J^{\text{(a)}}_{x}=J^{\text{(b)}}_{x}=J^{\text{(e)}}_{x}=J^{\text{(f)}}_{x}
\quad , \quad 
J^{\text{(c)}}_{x}=J^{\text{(d)}}_{x}=-8\lambda^{2}\,\text{Im}(t_{S})
\sum_{\bk}\left(\frac{u_{\bk}v_{\bk}}{E_{\bk}+U}\right)^{2},
\\
&J'^{\text{(a)}}_{x}=J'^{\text{(b)}}_{x}=J'^{\text{(e)}}_{x}=J'^{\text{(f)}}_{x}
\quad , \quad 
J'^{\text{(c)}}_{x}=J'^{\text{(d)}}_{x}=-8\lambda^{2}\,\text{Re}(t_{S})
\sum_{\bk}\left(\frac{u_{\bk}v_{\bk}}{E_{\bk}+U}\right)^{2}.
\end{split}
\end{equation}
Once we add the second order contribution, which corresponds to a conventional Josephson effect via the spin-flip tunnelling barrier, as well as the fourth order contribution, which corresponds to a parity-controlled $2\pi$-Josephson effect via the TRI TSC island \cite{bib:Schrade2018_SM}, we arrive at the effective Hamiltonian of Eq.~(10) in the main text,
\begin{equation}
\begin{split}
&H_{x,\text{even}}=-(J_{S}-J+\hat{x}\, J_{x})\cos\varphi \quad, \quad H_{x,\text{odd}}=-(J_{S}+J)\cos\varphi +  \hat{x} \, J'_{x}  \sin\varphi. 
\end{split}
\end{equation}
So far, we have again only considered Cooper pair transport sequences leading to terms in the effective Hamiltonian that depend on the SC phase difference. Terms which are independent of the SC phase difference do not affect the Josephson current and, for that reason, have been omitted for the $\hat{x}$-measurement protocol. However, those terms are clearly of relevance when pulsing the tunnel couplings to obtain a $\pi/8$-gate. As for the effective Hamiltonian for the $\hat{z}$-measurement protocol, we find that contributions that are independent of the SC phase difference and $\propto\hat{x}$ only occur for $n_{0}$ even. More concretely, the effective Hamiltonian modifies to,
\begin{equation}
\begin{split}
&H_{x,\text{even}}\rightarrow\hat{x}\tilde{J}_{x}-(J_{S}-J+\hat{x}\, J_{x})\cos\varphi \quad \text{with} \quad  
\tilde{J}_{x}=\tilde{J}^{\text{(a)}}_{x}+\tilde{J}^{\text{(b)}}_{x}+\tilde{J}^{\text{(c)}}_{x}+\tilde{J}^{\text{(d)}}_{x}+\tilde{J}^{\text{(e)}}_{x}+\tilde{J}^{\text{(f)}}_{x},
\end{split}
\end{equation}
where the additional coupling constants are given by,
\begin{equation}
\begin{split}
&\tilde{J}^{\text{(a)}}_{x}=\tilde{J}^{\text{(b)}}_{x}=\tilde{J}^{\text{(e)}}_{x}=\tilde{J}^{\text{(f)}}_{x}=4\lambda^{2}\,\text{Im}(t_{S})
\sum_{\bk}\frac{(u_{\bk}v_{\bk})^{2}}{(E_{\bk}+U)E_{\bk}}
\quad , \quad 
\tilde{J}^{\text{(c)}}_{x}=\tilde{J}^{\text{(d)}}_{x}=-8\lambda^{2}\,\text{Im}(t_{S})
\sum_{\bk}\left(\frac{u_{\bk}v_{\bk}}{E_{\bk}+U}\right)^{2}.
\end{split}
\end{equation}
Before closing this first section of the Supplemental Material, we note that the third-order contributions to $H_{z,\text{even}}$ and $H_{x,\text{even}}$ only depend on $\text{Im}(t_{N})$ and $\text{Im}(t_{S})$, respectively. In particular, spin-flip tunnelling contribution with $\text{Re}(t_{S})\neq0$ do not alter the form of $H_{z,\text{even}}$. Similarly, normal tunnelling contribution with $\text{Re}(t_{N})\neq0$ also do not alter the form of $H_{x,\text{even}}$. Hence, we are able to relax our initial assumptions from $t_{S}=0$ to $\text{Im}(t_{S})=0$ for $H_{z,\text{even}}$ and from $t_{N}=0$ to $\text{Im}(t_{N})=0$ for $H_{x,\text{even}}$. Similar arguments apply to $H_{z,\text{odd}}$ and $H_{x,\text{odd}}$. Here, our initial assumption $t_{S}=0$ relaxes to $\text{Re}(t_{S})=0$ for $H_{z,\text{odd}}$ and $t_{N}=0$ relaxes to $\text{Re}(t_{N})=0$ for $H_{x,\text{odd}}$.

\section{Effective Hamiltonian for the two-qubit gates}
In this second section of the Supplemental Material, we sketch the derivation of the effective Hamiltonian for the two-qubit gates as given in Eq.~(11) of the main text. Up to fourth order in the couplings, the general form of the effective Hamiltonian reads,
\begin{equation}
\begin{split}
H_{ab}&= 
- P_{ab} H_{T,ab} \left[H_{0,ab}^{-1}(1-P_{ab})H_{T,ab}\right]^{3}P_{ab}.
\end{split}
\end{equation}
\begin{figure}[!t] \centering
\includegraphics[width=0.9\linewidth] {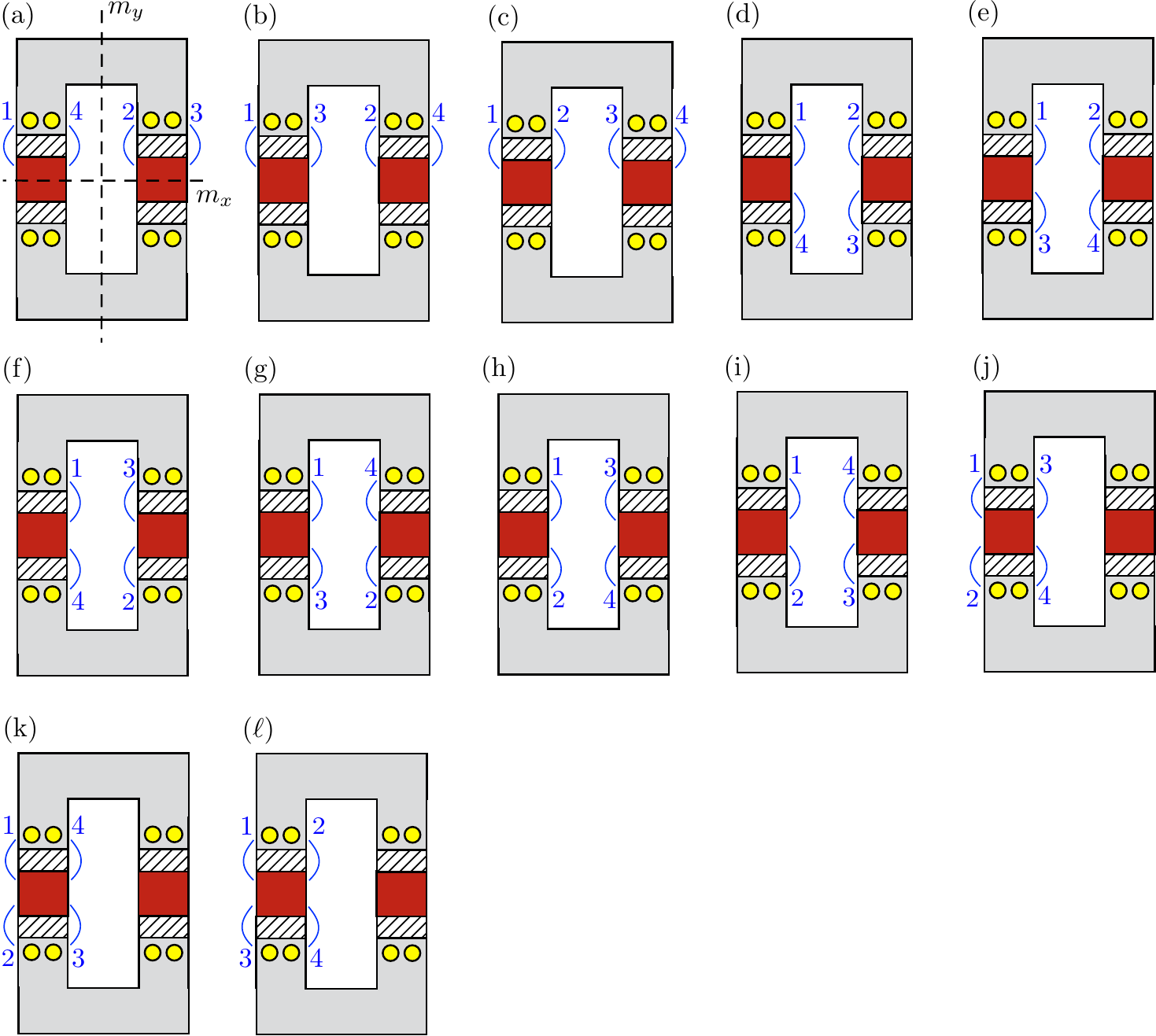}
\caption{(Color online)
Fourth-order sequences of intermediate states (up to hermitian-conjugation and mirror operations $m_{x}$, $m_{y}$) relevant for the two-qubit gate.
}\label{fig:2_SM}
\end{figure}
Here, we have dropped the second order contribution as it only leads to a constant shift in energy and consequently contributes neither to the entanglement generation between the two TRI TS islands nor the Josephson current between the SC grains. 
Moreover, $H_{0,ab}$ is the Hamiltonian of the uncoupled system comprised of the two SC leads and the two TRI TSC islands, $H_{T,ab}$ is the tunnelling Hamiltonian between the two SC leads and the TRI TSC islands and $P_{ab}$ is the projector on the reduced Hilbert space consisting of the BCS ground states of the SC leads as well as the charge ground states of both TRI TS islands. For simplicity, we will assume that both the TRI TSC islands as well as all their tunnel couplings to the SC leads are identical. In particular, both TRI TSC islands are tuned to a Coulomb valley with $n_{0}$ units of charge in the ground state and both are coupled to the ground through capacitors of equal capacitance $C$. The tunnelling amplitude between the TRI TSC islands and the SC leads will be denoted by $\lambda$. 

To evaluate the effective Hamiltonian, we first list the different types of sequences of intermediate states, see Fig.~\ref{fig:2_SM}. Then we compute the effective Hamiltonian for each type of sequence separately. A summation over all the different types then produces the final expression for the effective Hamiltonian, 
\begin{equation}
\begin{split}
H_{ab}
&=
H^{\text{(a)}}_{ab}
+
H^{\text{(b)}}_{ab}
+
H^{\text{(c)}}_{ab}
+
H^{\text{(d)}}_{ab}
+
H^{\text{(e)}}_{ab}
+
H^{\text{(f)}}_{ab}
\\
&+
H^{\text{(g)}}_{ab}
+
H^{\text{(h)}}_{ab}
+
H^{\text{(i)}}_{ab}
+
H^{\text{(j)}}_{ab}
+
H^{\text{(k)}}_{ab}
+
H^{\text{($\ell$)}}_{ab}
\end{split}
\end{equation}
Before going into the details of the derivation, we first display the full result for all types of contributions, 
\begin{equation}
\begin{split}
&H^{\text{(a)}}_{ab}
=
(-1)^{n_{0}+1}2J_{1}\cos\varphi \quad, \quad
H^{\text{(b)}}_{ab}
=
(-1)^{n_{0}+1}2J_{2}\cos\varphi  \quad, \quad
H^{\text{(c)}}_{ab}
=
(-1)^{n_{0}+1}2J_{3}\cos\varphi
\\
&H^{\text{(d)}}_{ab}
=
[(-1)^{n_{0}+1}J_{1}\cos\varphi-J_{4}](\hat{x}_{a}\hat{x}_{b}+\hat{z}_{a}\hat{z}_{b}) \quad, \quad
H^{\text{(e)}}_{ab}
=
[(-1)^{n_{0}+1}J_{2}\cos\varphi-J_{5}](\hat{x}_{a}\hat{x}_{b}+\hat{z}_{a}\hat{z}_{b})
\\
&H^{\text{(f)}}_{ab}
=
[(-1)^{n_{0}+1}J_{7}\cos\varphi+J_{6}](\hat{x}_{a}\hat{x}_{b}+\hat{z}_{a}\hat{z}_{b}) \quad\ , \quad
H^{\text{(g)}}_{ab}
=
[(-1)^{n_{0}+1}J_{9}\cos\varphi+J_{8}](\hat{x}_{a}\hat{x}_{b}+\hat{z}_{a}\hat{z}_{b})
\\
&
H^{\text{(h)}}_{ab}=
[(-1)^{n_{0}+1}J_{11}\cos\varphi-J_{10}](\hat{x}_{a}\hat{x}_{b}+\hat{z}_{a}\hat{z}_{b})\hspace{5pt} , \quad
H^{\text{(i)}}_{ab}
=
[(-1)^{n_{0}+1}J_{11}\cos\varphi+J_{12}](\hat{x}_{a}\hat{x}_{b}+\hat{z}_{a}\hat{z}_{b})
\\
&H^{\text{(j)}}_{ab}
=
-J_{13}\hat{y}_{a}\hat{y}_{b}
+J_{14}\hat{y}_{a}\hat{y}_{b} \quad, \quad
H^{\text{(k)}}_{ab}
=
J_{15}\hat{y}_{a}\hat{y}_{b}
+
J_{16}\hat{y}_{a}\hat{y}_{b} \quad, \quad
H^{\text{($\ell$)}}_{ab}
=
-J_{17}\hat{y}_{a}\hat{y}_{b}
-J_{18}\hat{y}_{a}\hat{y}_{b}.
\end{split}
\end{equation}
Here, we have introduced the coupling constants
\begin{equation}
\begin{split}
&J_{1}=16\lambda^{4}
\sum_{\bk,\bq}\frac{u_{\bk}v_{\bk}u_{\bq}v_{\bq}}{(E_{\bk}+U)^{2}(E_{\bk}+E_{\bq})}
\hspace{75pt}, \quad
J_{2}=16\lambda^{4}
\sum_{\bk,\bq}\frac{u_{\bk}v_{\bk}u_{\bq}v_{\bq}}{(E_{\bk}+U)(E_{\bk}+E_{\bq})(E_{\bq}+U)}
\\
&J_{3}=16\lambda^{4}
\sum_{\bk,\bq}\frac{u_{\bk}v_{\bk}u_{\bq}v_{\bq}}{(E_{\bk}+U)(4U)(E_{\bq}+U)}
\hspace{65pt}, \quad
J_{4}=8\lambda^{4}
\sum_{\bk,\bq}
\frac{
(u_{\bq}v_{\bk})^{2}+(v_{\bq}u_{\bk})^{2}
}
{
(E_{\bk}+U)^{2}(E_{\bk}+E_{\bq})
}
\\
&J_{5}=8\lambda^{4}
\sum_{\bk,\bq}
\frac{
(u_{\bq}v_{\bk})^{2}+(v_{\bq}u_{\bk})^{2}
}
{
(E_{\bq}+U)(E_{\bk}+E_{\bq})(E_{\bk}+U)
}
\hspace{44pt}, \quad
J_{6}=8\lambda^{4}
\sum_{\bk,\bq}
\frac{
(u_{\bk}u_{\bq})^{2}+(v_{\bk}v_{\bq})^{2}
}
{
(E_{\bk}+U)^{2}(E_{\bk}+E_{\bq}+2U)
}
\\
&J_{7}=16\lambda^{4}
\sum_{\bk,\bq}
\frac{
u_{\bk}u_{\bq}v_{\bk}v_{\bq}
}
{
(E_{\bk}+U)^{2}(E_{\bk}+E_{\bq}+2U)
}
\hspace{52pt},\quad
J_{8}=8\lambda^{4}
\sum_{\bk,\bq}
\frac{
(u_{\bk}u_{\bq})^{2}+(v_{\bk}v_{\bq})^{2}
}
{
(E_{\bk}+U)(E_{\bk}+E_{\bq}+2U)(E_{\bq}+U)
}
\\
&J_{9}=16\lambda^{4}
\sum_{\bk,\bq}
\frac{
u_{\bk}u_{\bq}v_{\bk}v_{\bq}
}
{
(E_{\bk}+U)(E_{\bk}+E_{\bq}+2U)(E_{\bq}+U)
}
\hspace{17pt},\quad
J_{10}=8\lambda^{4}
\sum_{\bk,\bq}
\frac{
(u_{\bq}v_{\bk})^{2}+(v_{\bq}u_{\bk})^{2}
}
{
(E_{\bk}+U)(2U)(E_{\bq}+U)
}
\\
&J_{11}=16\lambda^{4}
\sum_{\bk,\bq}
\frac{
u_{\bk}u_{\bq}v_{\bk}v_{\bq}
}
{
(E_{\bk}+U)(2U)(E_{\bq}+U)
}
\hspace{64pt},\quad
J_{12}=8\lambda^{4}
\sum_{\bk,\bq}
\frac{
(u_{\bk}u_{\bq})^{2}+(v_{\bk}v_{\bq})^{2}
}
{
(E_{\bk}+U)(2U)(E_{\bq}+U)
}
\\
&J_{13}=
8\lambda^{4}
\sum_{\bk,\bq}
\frac{
(u_{\bq}v_{\bk}-u_{\bk}v_{\bq})^{2}
}
{
(E_{\bk}+U)(2U)(E_{\bq}+U)
}
\hspace{68pt},\quad
J_{14}=
8\lambda^{4}
\sum_{\bk,\bq}
\frac{
(u_{\bk}u_{\bq}+v_{\bk}v_{\bq})^{2}
}
{
(E_{\bk}+U)^{2}(E_{\bk}+E_{\bq}+2U)
}
\\
&J_{15}=
8\lambda^{4}
\sum_{\bk,\bq}
\frac{
(u_{\bk}u_{\bq}+v_{\bk}v_{\bq})^{2}
}
{
(E_{\bq}+U)(2U)(E_{\bk}+U)
}
\hspace{68pt},\quad
J_{16}=
8\lambda^{4}
\sum_{\bk,\bq}
\frac{
(u_{\bk}u_{\bq}+v_{\bk}v_{\bq})^{2}
}
{
(E_{\bq}+U)(E_{\bk}+E_{\bq}+2U)(E_{\bk}+U)
}
\\
&J_{17}=8\lambda^{4}
\sum_{\bk,\bq}
\frac{
(u_{\bq}v_{\bk}-u_{\bk}v_{\bq})^{2}
}
{
(E_{\bq}+U)(E_{\bk}+E_{\bq})(E_{\bk}+U)
}
\hspace{43pt},\quad
J_{18}=8\lambda^{4}
\sum_{\bk,\bq}
\frac{
(u_{\bq}v_{\bk}-u_{\bk}v_{\bq})^{2}
}
{
(E_{\bk}+U)^{2}
(E_{\bk}+E_{\bq})
}.
\end{split}
\end{equation}
As a next step, we collect the different contributions. This simplifies the expression of the effective Hamiltonian to
\begin{equation}
\begin{split}
H_{ab}
= 
(-1)^{n_{0}+1}J_{0}\cos\varphi
+
[(-1)^{n_{0}+1}J'_{xz}\cos\varphi+J_{xz}](\hat{x}_{a}\hat{x}_{b}+\hat{z}_{a}\hat{z}_{b})
+
J_{y}\hat{y}_{a}\hat{y}_{b}
\end{split}
\end{equation}
with the coupling constants
\begin{equation}
\begin{split}
J_{0}&=J_{1}+J_{2}+J_{3}
\\
J'_{xz}&=J_{1}+J_{2}+J_{7}+J_{9}+2J_{11}
\\
J_{xz}&=-J_{4}-J_{5}+J_{6}+J_{8}-J_{10}+J_{12}
\\
J'_{y}&=J_{13}+J_{14}-J_{15}+J_{16}+J_{17}-J_{18}.
\end{split}
\end{equation}
Now that we have presented the full expression for the effective Hamiltonian we will give an overview of the derivation of the individual contributions. 

\subsection{Sequences of intermediate states corresponding to contributions $\boldsymbol{\propto J_{1},J_{2},J_{7},J_{9},J_{11}}$}
In this first subsection, we discuss sequences of intermediate states which lead to contributions $\propto J_{1},J_{2},J_{7},J_{9},J_{11}$ in the effective Hamiltonian. Initially, we will present two examples which yield contributions $\propto J_{1}$ in the effective Hamiltonian. Subsequently, we will explain how additional examples for the contributions $\propto J_{2},J_{7},J_{9},J_{11}$ can be obtained from these findings. Our first example is given by 
\begin{equation}
\begin{split}
&\quad\
P_{ab}(
\lambda
c^{\dag}_{\text{L},-\bk \da}
\gamma_{b,\text{L},\da}
e^{-i\phi_{2}/2}
)
(
\lambda
\gamma_{b,\text{R},\da}
c_{\text{R},-\bq \da}
e^{i\phi_{2}/2}
)
(
\lambda
\gamma_{a,\text{R},\ua}
c_{\text{R},\bq \ua}
e^{i\phi_{1}/2}
)
(
\lambda
c^{\dag}_{\text{L},\bk \ua}
\gamma_{a,\text{L},\ua}
e^{-i\phi_{1}/2}
)
P_{ab}
\\
&=
P_{ab}\lambda^{4}
(
c^{\dag}_{\text{L},-\bk \da}
\gamma_{b,\text{L},\da}
\gamma_{b,\text{R},\da}
c_{\text{R},-\bq \da}
\gamma_{a,\text{R},\ua}
c_{\text{R},\bq \ua}
c^{\dag}_{\text{L},\bk \ua}
\gamma_{a,\text{L},\ua}
)
P_{ab}
\\
&=
-
P_{ab}\lambda^{4}
(
\gamma_{a,\text{R},\ua}
\gamma_{a,\text{L},\ua}
\gamma_{b,\text{R},\da}
\gamma_{b,\text{L},\da}
)
(
c^{\dag}_{\text{L},-\bk \da}
c_{\text{R},-\bq \da}
c_{\text{R},\bq \ua}
c^{\dag}_{\text{L},\bk \ua}
)
P_{ab}
\\
&=
(-1)^{n_{0}+1}P_{ab}\lambda^{4}
\hat{z}_{a}\hat{z}_{b}
(
c^{\dag}_{\text{L},-\bk \da}
c_{\text{R},-\bq \da}
c_{\text{R},\bq \ua}
c^{\dag}_{\text{L},\bk \ua}
)
P_{ab}
\\
&=
(-1)^{n_{0}}
e^{i(\varphi_{\text{R}}-\varphi_{\text{L}})}
P_{ab}\lambda^{4}
v_{\bk}
u_{\bq}
v_{\bq}
u_{\bk}
\hat{z}_{a}\hat{z}_{b}
(
\gamma_{\text{L},\bk \ua}
\gamma_{\text{R},-\bq \da}
\gamma^{\dag}_{\text{R},-\bq \da}
\gamma^{\dag}_{\text{L},\bk \ua}
)
P_{ab}
\\
&=
(-1)^{n_{0}}
e^{i(\varphi_{\text{R}}-\varphi_{\text{L}})}
P_{ab}\lambda^{4}
v_{\bk}
u_{\bq}
v_{\bq}
u_{\bk}
\hat{z}_{a}\hat{z}_{b}
P_{ab}
\label{c1}
\end{split}
\end{equation}
and thus leads to a term $\propto\hat{z}_{a}\hat{z}_{b}$. In the second example the coupling to the MKPs is different compared to first example, 

\begin{equation}
\begin{split}
&\quad\
P_{ab}(
\lambda
c^{\dag}_{\text{L},-\bk \da}
\gamma_{b,\text{L},\da}
e^{-i\phi_{2}/2}
)
(
\lambda
\gamma_{b,\text{R},\ua}
c_{\text{R},\bq \ua}
e^{i\phi_{2}/2}
)
(
\lambda
\gamma_{a,\text{R},\da}
c_{\text{R},-\bq \da}
e^{i\phi_{1}/2}
)
(
\lambda
c^{\dag}_{\text{L},\bk \ua}
\gamma_{a,\text{L},\ua}
e^{-i\phi_{1}/2}
)
P_{ab}
\\
&=
P_{ab}\lambda^{4}
(
c^{\dag}_{\text{L},-\bk \da}
\gamma_{b,\text{L},\da}
\gamma_{b,\text{R},\ua}
c_{\text{R},\bq \ua}
\gamma_{a,\text{R},\da}
c_{\text{R},-\bq \da}
c^{\dag}_{\text{L},\bk \ua}
\gamma_{a,\text{L},\ua}
)
P_{ab}
\\
&=
P_{ab}\lambda^{4}
(
\gamma_{b,\text{R},\ua}
\gamma_{b,\text{L},\da}
\gamma_{a,\text{L},\ua}
\gamma_{a,\text{R},\da}
)
(
c^{\dag}_{\text{L},-\bk \da}
c_{\text{R},\bq \ua}
c_{\text{R},-\bq \da}
c^{\dag}_{\text{L},\bk \ua}
)
P_{ab}
\\
&=
(-1)^{n_{0}}
P_{ab}\lambda^{4}
\hat{x}_{a}\hat{x}_{b}
(
c^{\dag}_{\text{L},-\bk \da}
c_{\text{R},\bq \ua}
c_{\text{R},-\bq \da}
c^{\dag}_{\text{L},\bk \ua}
)
P_{ab}
\\
&=
(-1)^{n_{0}}
e^{i(\varphi_{\text{R}}-\varphi_{\text{L}})}
P_{ab}\lambda^{4}
v_{\bk}
u_{\bq}
v_{\bq}
u_{\bk}
\hat{x}_{a}\hat{x}_{b}
(
\gamma_{\text{L},\bk \ua}
\gamma_{\text{R},\bq \ua}
\gamma^{\dag}_{\text{R},\bq \ua}
\gamma^{\dag}_{\text{L},\bk \ua}
)
P_{ab}
\\
&=
(-1)^{n_{0}}
e^{i(\varphi_{\text{R}}-\varphi_{\text{L}})}
P_{ab}\lambda^{4}
v_{\bk}
u_{\bq}
v_{\bq}
u_{\bk}
\hat{x}_{a}\hat{x}_{b}
P_{ab}
\label{c2}
\end{split}
\end{equation}
which leads to a term $\propto\hat{x}_{a}\hat{x}_{b}$. The energy denominator for both examples is
given by $-1/[(E_{\bk}+U)^{2}(E_{\bk}+E_{\bq})]$. Hence, after combining the above results with the corresponding hermitian-conjugated sequences and summing over all momenta, we indeed find a contribution $\propto J_{1}$.

Finally, we point out that examples for contributions $\propto J_{2},J_{7},J_{9},J_{11}$ can be obtained by suitably commuting the terms in the first line of Eq.~\eqref{c1} and Eq.~\eqref{c2}. The different coupling constants arise because the energy denominators for the resulting processes will be different than the ones we used for the two examples given above. 

\subsection{Sequences of intermediate states corresponding to contributions $\boldsymbol{\propto J_{4},J_{5},J_{6},J_{8},J_{10},J_{12}}$}
In this second subsection, we continue our overview on the sequences of intermediate states which contribute to the effective Hamiltonian. More specifically, we will examine sequences that give contributions $\propto J_{4},J_{5},J_{6},J_{8},J_{10},J_{12}$. As a first step, we introduce two examples which produce contributions $\propto J_{4}$. The first example is given by
\begin{equation}
\begin{split}
&\quad\
P_{ab}(
\lambda
\gamma_{b,\text{L},\ua}
c_{\text{L},\bk \ua}
e^{i\phi_{2}/2}
)
(
\lambda
c^{\dag}_{\text{R},\bq \ua}
\gamma_{b,\text{R},\ua}
e^{-i\phi_{2}/2}
)
(
\lambda
\gamma_{a,\text{R},\ua}
c_{\text{R},\bq \ua}
e^{i\phi_{1}/2}
)
(
\lambda
c^{\dag}_{\text{L},\bk \ua}
\gamma_{a,\text{L},\ua}
e^{-i\phi_{1}/2}
)
P_{ab}
\\
&=
P_{ab}\lambda^{4}
(
\gamma_{b,\text{L},\ua}
c_{\text{L},\bk \ua}
c^{\dag}_{\text{R},\bq \ua}
\gamma_{b,\text{R},\ua}
\gamma_{a,\text{R},\ua}
c_{\text{R},\bq \ua}
c^{\dag}_{\text{L},\bk \ua}
\gamma_{a,\text{L},\ua}
)
P_{ab}
\\
&=
-
P_{ab}\lambda^{4}
(
\gamma_{a,\text{R},\ua}
\gamma_{a,\text{L},\ua}
\gamma_{b,\text{R},\ua}
\gamma_{b,\text{L},\ua}
)
(
c_{\text{L},\bk \ua}
c^{\dag}_{\text{R},\bq \ua}
c_{\text{R},\bq \ua}
c^{\dag}_{\text{L},\bk \ua}
)
P_{ab}
\\
&=
P_{ab}\lambda^{4}
\hat{z}_{a}\hat{z}_{b}
(
c_{\text{L},\bk \ua}
c^{\dag}_{\text{R},\bq \ua}
c_{\text{R},\bq \ua}
c^{\dag}_{\text{L},\bk \ua}
)
P_{ab}
\\
&=
P_{ab}\lambda^{4}
\hat{z}_{a}\hat{z}_{b}
(
v_{\bq}
u_{\bk}
)^{2}
(
\gamma_{\text{L},\bk \ua}
\gamma_{\text{R},-\bq \da}
\gamma^{\dag}_{\text{R},-\bq \da}
\gamma^{\dag}_{\text{L},\bk \ua}
)
P_{ab}
\\
&=
P_{ab}\lambda^{4}
\hat{z}_{a}\hat{z}_{b}
(
v_{\bq}
u_{\bk}
)^{2}
P_{ab}
\label{c3}
\end{split}
\end{equation}
and it leads to a term $\propto\hat{z}_{a}\hat{z}_{b}$. The second example is given by 
\begin{equation}
\begin{split}
&\quad\
P_{ab}(
\lambda
\gamma_{b,\text{L},\ua}
c_{\text{L},\bk \ua}
e^{i\phi_{2}/2}
)
(
\lambda
c^{\dag}_{\text{R},\bq \da}
\gamma_{b,\text{R},\da}
e^{-i\phi_{2}/2}
)
(
\lambda
\gamma_{a,\text{R},\da}
c_{\text{R},\bq \da}
e^{i\phi_{1}/2}
)
(
\lambda
c^{\dag}_{\text{L},\bk \ua}
\gamma_{a,\text{L},\ua}
e^{-i\phi_{1}/2}
)
P_{ab}
\\
&=
P_{ab}\lambda^{4}
(
\gamma_{b,\text{L},\ua}
c_{\text{L},\bk \ua}
c^{\dag}_{\text{R},\bq \da}
\gamma_{b,\text{R},\da}
\gamma_{a,\text{R},\da}
c_{\text{R},\bq \da}
c^{\dag}_{\text{L},\bk \ua}
\gamma_{a,\text{L},\ua}
)
P_{ab}
\\
&=
-
P_{ab}\lambda^{4}
(
\gamma_{a,\text{L},\ua}
\gamma_{a,\text{R},\da}
\gamma_{b,\text{L},\ua}
\gamma_{b,\text{R},\da}
)
(
c_{\text{L},\bk \ua}
c^{\dag}_{\text{R},\bq \da}
c_{\text{R},\bq \da}
c^{\dag}_{\text{L},\bk \ua}
)
P_{ab}
\\
&=
P_{ab}\lambda^{4}
\hat{x}_{a}\hat{x}_{b}
(
c_{\text{L},\bk \ua}
c^{\dag}_{\text{R},\bq \da}
c_{\text{R},\bq \da}
c^{\dag}_{\text{L},\bk \ua}
)
P_{ab}
\\
&=
P_{ab}\lambda^{4}
\hat{x}_{a}\hat{x}_{b}
(
u_{\bk}
v_{\bq}
)^{2}
(
\gamma_{\text{L},\bk \ua}
\gamma_{\text{R},-\bq \ua}
\gamma^{\dag}_{\text{R},-\bq \ua}
\gamma^{\dag}_{\text{L},\bk \ua}
)
P_{ab}
\\
&=
P_{ab}\lambda^{4}
\hat{x}_{a}\hat{x}_{b}
(
u_{\bk}
v_{\bq}
)^{2}
P_{ab},
\label{c4}
\end{split}
\end{equation}
and gives a term $\propto\hat{x}_{a}\hat{x}_{b}$. The energy denominator for both examples is given by $-1/[(E_{\bk}+U)^{2}(E_{\bk}+E_{\bq})]$. After carrying out the summation over all momenta, we see that both examples indeed contribute to the term $\propto J_{4}$ in the effective Hamiltonian. 
Moreover, we remark that examples for sequences of intermediate states that give contributions $\propto J_{5}$ and $\propto J_{10}$ can be obtained by appropriately commuting the terms in the round brackets in the first line of Eqs.~\eqref{c3} and \eqref{c4}. The required energy denominator for the examples $\propto J_{5}$ is $-1/[(E_{\bq}+U)(E_{\bk}+E_{\bq})(E_{\bk}+U)]$ and for examples $\propto J_{10}$ it is $-1/[(E_{\bq}+U)(2U)(E_{\bk}+U)]$.

We now proceed by presenting two examples of sequences of intermediate states leading to contributions $\propto J_{6}$. The first example is given by 
\begin{equation}
\begin{split}
&\quad\
P_{ab}(
\lambda
\gamma_{b,\text{L},\ua}
c_{\text{L},\bk \ua}
e^{i\phi_{2}/2}
)
(
\lambda
\gamma_{a,\text{R},\ua}
c_{\text{R},\bq \ua}
e^{i\phi_{1}/2}
)
(
\lambda
c^{\dag}_{\text{R},\bq \ua}
\gamma_{b,\text{R},\ua}
e^{-i\phi_{2}/2}
)
(
\lambda
c^{\dag}_{\text{L},\bk \ua}
\gamma_{a,\text{L},\ua}
e^{-i\phi_{1}/2}
)
P_{ab}
\\
&=
P_{ab}\lambda^{4}
(
\gamma_{b,\text{L},\ua}
c_{\text{L},\bk \ua}
\gamma_{a,\text{R},\ua}
c_{\text{R},\bq \ua}
c^{\dag}_{\text{R},\bq \ua}
\gamma_{b,\text{R},\ua}
c^{\dag}_{\text{L},\bk \ua}
\gamma_{a,\text{L},\ua}
)
P_{ab}
\\
&=
P_{ab}\lambda^{4}
(
\gamma_{a,\text{R},\ua}
\gamma_{a,\text{L},\ua}
\gamma_{b,\text{R},\ua}
\gamma_{b,\text{L},\ua}
)
(
c_{\text{L},\bk \ua}
c_{\text{R},\bq \ua}
c^{\dag}_{\text{R},\bq \ua}
c^{\dag}_{\text{L},\bk \ua}
)
P_{ab}
\\
&=
-
P_{ab}\lambda^{4}
\hat{z}_{a}\hat{z}_{b}
(
c_{\text{L},\bk \ua}
c_{\text{R},\bq \ua}
c^{\dag}_{\text{R},\bq \ua}
c^{\dag}_{\text{L},\bk \ua}
)
P_{ab}
\\
&=
-
P_{ab}\lambda^{4}
\hat{z}_{a}\hat{z}_{b}
(
u_{\bq}
u_{\bk}
)^{2}
(
\gamma_{\text{L},\bq \ua}
\gamma_{\text{R},\bq \ua}
\gamma^{\dag}_{\text{R},\bq \ua}
\gamma^{\dag}_{\text{L},\bk \ua}
)
P_{ab}
\\
&=
-
P_{ab}\lambda^{4}
\hat{z}_{a}\hat{z}_{b}
(
u_{\bq}
u_{\bk}
)^{2}
P_{ab}
\label{c5}
\end{split}
\end{equation}
leading to a term $\propto\hat{z}_{a}\hat{z}_{b}$ and the second example is given  
\begin{equation}
\begin{split}
&\quad\
P_{ab}(
\lambda
\gamma_{b,\text{L},\ua}
c_{\text{L},\bk \ua}
e^{i\phi_{2}/2}
)
(
\lambda
\gamma_{a,\text{R},\da}
c_{\text{R},\bq \da}
e^{i\phi_{1}/2}
)
(
\lambda
c^{\dag}_{\text{R},\bq \da}
\gamma_{b,\text{R},\da}
e^{-i\phi_{2}/2}
)
(
\lambda
c^{\dag}_{\text{L},\bk \ua}
\gamma_{a,\text{L},\ua}
e^{-i\phi_{1}/2}
)
P_{ab}
\\
&=
P_{ab}\lambda^{4}
(
\gamma_{b,\text{L},\ua}
c_{\text{L},\bk \ua}
\gamma_{a,\text{R},\da}
c_{\text{R},\bq \da}
c^{\dag}_{\text{R},\bq \da}
\gamma_{b,\text{R},\da}
c^{\dag}_{\text{L},\bk \ua}
\gamma_{a,\text{L},\ua}
)
P_{ab}
\\
&=
P_{ab}\lambda^{4}
(
\gamma_{a,\text{L},\ua}
\gamma_{a,\text{R},\da}
\gamma_{b,\text{L},\ua}
\gamma_{b,\text{R},\da}
)
(
c_{\text{L},\bk \ua}
c_{\text{R},\bq \da}
c^{\dag}_{\text{R},\bq \da}
c^{\dag}_{\text{L},\bk \ua}
)
P_{ab}
\\
&=
-
P_{ab}\lambda^{4}
\hat{x}_{a}\hat{x}_{b}
(
c_{\text{L},\bk \ua}
c_{\text{R},\bq \da}
c^{\dag}_{\text{R},\bq \da}
c^{\dag}_{\text{L},\bk \ua}
)
P_{ab}
\\
&=
-
P_{ab}\lambda^{4}
\hat{x}_{a}\hat{x}_{b}
(
u_{\bk}
u_{\bq}
)^{2}
(
\gamma_{\text{L},\bk \ua}
\gamma_{\text{L},\bq \da}
\gamma^{\dag}_{\text{L},\bq \da}
\gamma^{\dag}_{\text{L},\bk \ua}
)
P_{ab}
\\
&=
-
P_{ab}\lambda^{4}
\hat{x}_{a}\hat{x}_{b}
(
u_{\bk}
u_{\bq}
)^{2}
P_{ab}
\label{c6}
\end{split}
\end{equation}
producing a term $\propto\hat{x}_{a}\hat{x}_{b}$. The energy denominator for both examples is given by $-1/[(E_{\bk}+U)^{2}(E_{\bk}+E_{\bq}+2U)]$. Hence, once we have summed over all momenta, we conclude that both examples give contributions $\propto J_{6}$. Finally, we remark that examples for contributions $\propto J_{8}$ and $\propto J_{12}$ can be obtained by appropriately commuting the terms in the round brackets in the first line of Eqs.~\eqref{c5} and \eqref{c6}. The only difference occurs in the energy denominator. The latter is given by $-1/[(E_{\bq}+U)(E_{\bk}+E_{\bq}+2U)(E_{\bk}+U)]$ for the contributions $\propto J_{8}$ and by $-1/[(E_{\bq}+U)(2U)(E_{\bk}+U)]$ for the contributions $\propto J_{12}$.

\subsection{Sequences of intermediate states corresponding to contributions $\boldsymbol{\propto J_{13},J_{15}}$}
In this third subsection we discuss sequences of intermediate states that lead to contributions $\propto J_{13}, J_{14}$ in the effective Hamiltonian. An examples that lead to a contribution $\propto J_{13}$ is given by 
\begin{equation}
\begin{split}
&\quad\
P_{ab}(
\lambda
c^{\dag}_{\text{L},\bq \da}
\gamma_{b,\text{L},\da}
e^{-i\phi_{2}/2}
)
(
\lambda
\gamma_{a,\text{L},\da}
c_{\text{L},\bq \da}
e^{i\phi_{1}/2}
)
(
\lambda
\gamma_{b,\text{L},\ua}
c_{\text{L},\bk \ua}
e^{i\phi_{2}/2}
)
(
\lambda
c^{\dag}_{\text{L},\bk \ua}
\gamma_{a,\text{L},\ua}
e^{-i\phi_{1}/2}
)
P_{ab}
\\
&=
P_{ab}\lambda^{4}
(
c^{\dag}_{\text{L},\bq \da}
\gamma_{b,\text{L},\da}
\gamma_{a,\text{L},\da}
c_{\text{L},\bq \da}
\gamma_{b,\text{L},\ua}
c_{\text{L},\bk \ua}
c^{\dag}_{\text{L},\bk \ua}
\gamma_{a,\text{L},\ua}
)
P_{ab}
\\
&=
-
P_{ab}\lambda^{4}
(
\gamma_{a,\text{L},\ua}
\gamma_{a,\text{L},\da}
\gamma_{b,\text{L},\ua}
\gamma_{b,\text{L},\da}
)
(
c^{\dag}_{\text{L},\bq \da}
c_{\text{L},\bq \da}
c_{\text{L},\bk \ua}
c^{\dag}_{\text{L},\bk \ua}
)
P_{ab}
\\
&=
P_{ab}\lambda^{4}
\hat{y}_{a}\hat{y}_{b}
(
c^{\dag}_{\text{L},\bq \da}
c_{\text{L},\bq \da}
c_{\text{L},\bk \ua}
c^{\dag}_{\text{L},\bk \ua}
)
P_{ab}
\\
&=
P_{ab}\lambda^{4}
(
v_{\bq}
u_{\bk}
)^{2}
\hat{y}_{a}\hat{y}_{b}
(
\gamma_{\text{L},-\bq \ua}
\gamma^{\dag}_{\text{L},-\bq \ua}
\gamma_{\text{L},\bk \ua}
\gamma^{\dag}_{\text{L},\bk \ua}
)
P_{ab}
\\
&=
P_{ab}\lambda^{4}
(
v_{\bq}
u_{\bk}
)^{2}
\hat{y}_{a}\hat{y}_{b}
P_{ab}.
\label{c9}
\end{split}
\end{equation}
and gives a term $\propto\hat{y}_{a}\hat{y}_{b}$.
The energy denominator for thhis sequences is given by $-1/[(E_{\bq}+U)(2U)(E_{\bk}+U)]$. Hence, after summing over all momenta we verify that both sequences indeed contribute to the term $\propto J_{13}$ in the effective Hamiltonian.

Providing examples of sequences that give contributions  $\propto J_{15}$ in the effective Hamiltonian requires
us to swap the first two terms in the round brackets in the first line of Eq.~\eqref{c9}. 
However, these terms do not commute. Hence, we need to re-evaluate the modified sequences. We find that,
\begin{equation}
\begin{split}
&\quad\
P_{ab}(
\lambda
\gamma_{a,\text{L},\da}
c_{\text{L},\bq \da}
e^{i\phi_{1}/2}
)
(
\lambda
c^{\dag}_{\text{L},\bq \da}
\gamma_{b,\text{L},\da}
e^{-i\phi_{2}/2}
)
(
\lambda
\gamma_{b,\text{L},\ua}
c_{\text{L},\bk \ua}
e^{i\phi_{2}/2}
)
(
\lambda
c^{\dag}_{\text{L},\bk \ua}
\gamma_{a,\text{L},\ua}
e^{-i\phi_{1}/2}
)
P_{ab}
\\
&=
P_{ab}\lambda^{4}
(
\gamma_{a,\text{L},\da}
c_{\text{L},\bq \da}
c^{\dag}_{\text{L},\bq \da}
\gamma_{b,\text{L},\da}
\gamma_{b,\text{L},\ua}
c_{\text{L},\bk \ua}
c^{\dag}_{\text{L},\bk \ua}
\gamma_{a,\text{L},\ua}
)
P_{ab}
\\
&=
P_{ab}\lambda^{4}
(
\gamma_{a,\text{L},\ua}
\gamma_{a,\text{L},\da}
\gamma_{b,\text{L},\ua}
\gamma_{b,\text{L},\da}
)
(
c_{\text{L},\bq \da}
c^{\dag}_{\text{L},\bq \da}
c_{\text{L},\bk \ua}
c^{\dag}_{\text{L},\bk \ua}
)
P_{ab}
\\
&=
-
P_{ab}\lambda^{4}
\hat{y}_{a}\hat{y}_{b}
(
c_{\text{L},\bq \da}
c^{\dag}_{\text{L},\bq \da}
c_{\text{L},\bk \ua}
c^{\dag}_{\text{L},\bk \ua}
)
P_{ab}
\\
&=
-
P_{ab}\lambda^{4}
(
u_{\bq}
u_{\bk}
)^{2}
\hat{y}_{a}\hat{y}_{b}
(
\gamma_{\text{L},\bq \da}
\gamma^{\dag}_{\text{L},\bq \da}
\gamma_{\text{L},\bk \ua}
\gamma^{\dag}_{\text{L},\bk \ua}
)
P_{ab}
\\
&=
-
P_{ab}\lambda^{4}
(
u_{\bq}
u_{\bk}
)^{2}
\hat{y}_{a}\hat{y}_{b}
P_{ab}.
\label{c11}
\end{split}
\end{equation}
The energy denominator for this sequence is still given by $-1/[(E_{\bq}+U)(2U)(E_{\bk}+U)]$. Consequently, after summing over all momenta, we recognize that the example given in Eq.~\eqref{c11} contributes to the term $\propto J_{15}$ in the effective Hamiltonian.

\subsection{Sequences of intermediate states corresponding to contributions $\boldsymbol{\propto J_{14},J_{16}}$}
In this fourth subsection, we discuss sequences of intermediate states that contribute to the terms $\propto J_{14},J_{16}$ in the effective Hamiltonian. We begin by presenting an examples fo a sequence of intermediate states that yield contribution $\propto J_{14}$ in the effective Hamiltonian, 
\begin{equation}
\begin{split}
&\quad\
P_{ab}(
\lambda
c^{\dag}_{\text{L},-\bk \da}
\gamma_{b,\text{L},\da}
e^{-i\phi_{2}/2}
)
(
\lambda
\gamma_{a,\text{L},\da}
c_{\text{L},-\bq \da}
e^{i\phi_{1}/2}
)
(
\lambda
\gamma_{b,\text{L},\ua}
c_{\text{L},\bq \ua}
e^{i\phi_{2}/2}
)
(
\lambda
c^{\dag}_{\text{L},\bk \ua}
\gamma_{a,\text{L},\ua}
e^{-i\phi_{1}/2}
)
P_{ab}
\\
&=
P_{ab}\lambda^{4}
(
c^{\dag}_{\text{L},-\bk \da}
\gamma_{b,\text{L},\da}
\gamma_{a,\text{L},\da}
c_{\text{L},-\bq \da}
\gamma_{b,\text{L},\ua}
c_{\text{L},\bq \ua}
c^{\dag}_{\text{L},\bk \ua}
\gamma_{a,\text{L},\ua}
)
P_{ab}
\\
&=
-
P_{ab}\lambda^{4}
(
\gamma_{a,\text{L},\ua}
\gamma_{a,\text{L},\da}
\gamma_{b,\text{L},\ua}
\gamma_{b,\text{L},\da}
)
(
c^{\dag}_{\text{L},-\bk \da}
c_{\text{L},-\bq \da}
c_{\text{L},\bq \ua}
c^{\dag}_{\text{L},\bk \ua}
)
P_{ab}
\\
&=
P_{ab}\lambda^{4}
\hat{y}_{a}\hat{y}_{b}
(
c^{\dag}_{\text{L},-\bk \da}
c_{\text{L},-\bq \da}
c_{\text{L},\bq \ua}
c^{\dag}_{\text{L},\bk \ua}
)
P_{ab}
\\
&=
-
P_{ab}\lambda^{4}
v_{\bk}
u_{\bq}
v_{\bq}
u_{\bk}
\hat{y}_{a}\hat{y}_{b}
(
\gamma_{\text{L},\bk \ua}
\gamma_{\text{L},-\bq \da}
\gamma^{\dag}_{\text{L},-\bq \da}
\gamma^{\dag}_{\text{L},\bk \ua}
)
P_{ab}
\\
&=
-
P_{ab}\lambda^{4}
v_{\bk}
u_{\bq}
v_{\bq}
u_{\bk}
\hat{y}_{a}\hat{y}_{b}
P_{ab}
\label{c14}
\end{split}
\end{equation}
The energy denominator for this example is given by $-1/[(E_{\bk}+U)^{2}(E_{\bk}+E_{\bq}+2U)]$. This means that after summing over all momenta, both sequences indeed
contribute to the term $\propto J_{14}$ in the effective Hamiltonian.

Finally, we note that examples for sequences of intermediate states leading to contributions $\propto J_{16}$ can be obtained by commuting the first two terms in the round brackets in the first line of Eq.~\eqref{c14} and adapting the energy denominators accordingly.

\subsection{Sequences of intermediate states corresponding to contributions $\boldsymbol{\propto J_{17},J_{18}}$}
In this final subsection, we give examples on sequences of intermediate states which lead to contributions $\propto J_{17}, J_{18}$ in the effective Hamiltonian. An example leading to a contribution $\propto J_{18}$ is given by
\begin{equation}
\begin{split}
&\quad\
P_{ab}(
\lambda
c^{\dag}_{\text{L},-\bk \da}
\gamma_{b,\text{L},\da}
e^{-i\phi_{2}/2}
)
(
\lambda
\gamma_{b,\text{L},\ua}
c_{\text{L},\bq \ua}
e^{i\phi_{2}/2}
)
(
\lambda
\gamma_{a,\text{L},\da}
c_{\text{L},-\bq \da}
e^{i\phi_{1}/2}
)
(
\lambda
c^{\dag}_{\text{L},\bk \ua}
\gamma_{a,\text{L},\ua}
e^{-i\phi_{1}/2}
)
P_{ab}
\\
&=
P_{ab}\lambda^{4}
(
c^{\dag}_{\text{L},-\bk \da}
\gamma_{b,\text{L},\da}
\gamma_{b,\text{L},\ua}
c_{\text{L},\bq \ua}
\gamma_{a,\text{L},\da}
c_{\text{L},-\bq \da}
c^{\dag}_{\text{L},\bk \ua}
\gamma_{a,\text{L},\ua}
)
P_{ab}
\\
&=
P_{ab}\lambda^{4}
(
\gamma_{a,\text{L},\ua}
\gamma_{a,\text{L},\da}
\gamma_{b,\text{L},\ua}
\gamma_{b,\text{L},\da}
)
(
c^{\dag}_{\text{L},-\bk \da}
c_{\text{L},\bq \ua}
c_{\text{L},-\bq \da}
c^{\dag}_{\text{L},\bk \ua}
)
P_{ab}
\\
&=
-
P_{ab}\lambda^{4}
\hat{y}_{a}\hat{y}_{b}
(
c^{\dag}_{\text{L},-\bk \da}
c_{\text{L},\bq \ua}
c_{\text{L},-\bq \da}
c^{\dag}_{\text{L},\bk \ua}
)
P_{ab}
\\
&=
-
P_{ab}\lambda^{4}
v_{\bk}
u_{\bq}
v_{\bq}
u_{\bk}
\hat{y}_{a}\hat{y}_{b}
(
\gamma_{\text{L},\bk \ua}
\gamma_{\text{L},\bq \ua}
\gamma^{\dag}_{\text{L},\bq \ua}
\gamma^{\dag}_{\text{L},\bk \ua}
)
P_{ab}
\\
&=
-
P_{ab}\lambda^{4}
v_{\bk}
u_{\bq}
v_{\bq}
u_{\bk}
\hat{y}_{a}\hat{y}_{b}
P_{ab}.
\label{c16}
\end{split}
\end{equation}
The energy denominator is given by $-1/[(E_{\bk}+U)^{2}(E_{\bk}+E_{\bq})]$. 
Hence, after summing over all momenta, we find a contribution $\propto J_{18}$.  

An example for a sequence of intermediate states that yields a contribution $\propto J_{17}$ 
can be obtained by swapping the first two terms in round brackets in the first line of Eq.~\eqref{c16}.
More concretely, 
\begin{equation}
\begin{split}
&\quad\
P_{ab}(
\lambda
\gamma_{b,\text{L},\ua}
c_{\text{L},\bq \ua}
e^{i\phi_{2}/2}
)
(
\lambda
c^{\dag}_{\text{L},-\bk \da}
\gamma_{b,\text{L},\da}
e^{-i\phi_{2}/2}
)
(
\lambda
\gamma_{a,\text{L},\da}
c_{\text{L},-\bq \da}
e^{i\phi_{1}/2}
)
(
\lambda
c^{\dag}_{\text{L},\bk \ua}
\gamma_{a,\text{L},\ua}
e^{-i\phi_{1}/2}
)
P_{ab}
\\
&=
P_{ab}(
\lambda
c^{\dag}_{\text{L},-\bk \da}
\gamma_{b,\text{L},\da}
e^{-i\phi_{2}/2}
)
(
\lambda
\gamma_{b,\text{L},\ua}
c_{\text{L},\bq \ua}
e^{i\phi_{2}/2}
)
(
\lambda
\gamma_{a,\text{L},\da}
c_{\text{L},-\bq \da}
e^{i\phi_{1}/2}
)
(
\lambda
c^{\dag}_{\text{L},\bk \ua}
\gamma_{a,\text{L},\ua}
e^{-i\phi_{1}/2}
)
P_{ab}
\\
&=
-
P_{ab}\lambda^{4}
v_{\bk}
u_{\bq}
v_{\bq}
u_{\bk}
\hat{y}_{a}\hat{y}_{b}
P_{ab}.
\end{split}
\end{equation}
The energy denominator is now given by $-1/[(E_{\bq}+U)(E_{\bk}+E_{\bq})(E_{\bk}+U)]$. Consequently, after summing over all momenta, we find a contribution $\propto J_{17}$ in the effective Hamiltonian.
\\
\\
In summary, in this second section of the Supplemental Material, we have provided an extensive overview of the numerous contributions which make up the effective Hamiltonian for the two-qubit gate as given in Eq.~(11) of the main text.

\end{widetext}

\end{document}